\documentstyle[aps,epsfig,preprint]{revtex}

\def\upleftarrow#1{\overleftarrow{#1}}   
\def\uprightarrow#1{\overrightarrow{#1}}
\def\thru#1{\mathrel{\mathop{#1\!\!\!/}}}

\begin{document}
\tighten
\draft
\preprint{
\vbox{
\hbox{February 1997}
\hbox{DOE/ER/40762-114}
\hbox{U.MD. PP\# 97-085}
\hbox{INPP-UVA-97-01}
}}

\title{A STUDY OF OFF-FORWARD PARTON DISTRIBUTIONS}

\author{Xiangdong Ji\footnote{
	On leave of absence from Department of Physics,
	Massachussetts Institute of Technology,
	Cambridge, MA 02139} and W. Melnitchouk}
\address{Department of Physics,
	University of Maryland,
	College Park, MD 20742}
\author{X. Song}
\address{Department of Physics,
	University of Virginia,
	Charlottesville, VA 22901}

\maketitle

\begin{abstract}
An extensive theoretical analysis of off-forward parton distributions (OFPDs)
is presented. The OFPDs and the form factors of the quark energy-momentum
tensor are estimated at a low energy scale using a bag model. Relations among
the second moments of OFPDs, the form factors, and the fraction of the nucleon
spin carried by quarks are discussed. 
\end{abstract}

\pacs{PACS numbers: 12.38.Aw, 12.39.Ba, 13.40.Gp, 13.60.Hb}

\narrowtext

\section{Introduction}

One of the most important frontiers in strong interaction physics is
the study of the structure of the nucleon.
Despite considerable experimental and theoretical progress made over
the last forty years, there are still many unanswered questions.
An example is the intensive debate which has continued over the spin
structure of the nucleon, ever since the European Muon
Collaboration (EMC) published their initial data on the spin structure
function $g_1$ \cite{EMC}.
Traditionally, two types of observables related to the nucleon
structure have been studied mostly extensively: 
elastic form factors and parton (quark and gluon) distributions.
Electromagnetic form factors of the proton were first measured in
the mid 1950s \cite{FFP}, and in recent years measurements of those
of the neutron have been attempted and more are planned \cite{FFN}.
Weak form factors are also being measured through parity-violating
electron and neutrino scattering \cite{WEAK}.
On the other hand, the unpolarized quark and gluon distributions
have been systematically probed in deep-inelastic scattering and 
Drell-Yan processes since the discovery of quarks at SLAC in the
late 1960s.
The polarized quark distributions have also been studied in a number
of experiments in recent years \cite{SPINEXP}, and more data on these
are anticipated in the future from CERN, SLAC, HERA and RHIC.

In this paper, we present a first detailed study of a new type of nucleon
observable: the off-forward parton distribution (OFPD).
The OFPDs generalize and interpolate between the
ordinary parton distributions, measured for instance 
in deep-inelastic scattering, and the elastic
form factors, and therefore contain rich structural information. 
There are no data so far on these distributions, and a relatively
short theoretical history. To our knowledge, the OFPDs have 
come up independently in three different theoretical studies.
In the late 1980s, Geyer and collaborators \cite{GEYER} studied the
relation between the Altarelli-Parisi evolution for parton distributions
and the Brodsky-Lepage evolution for leading-twist meson wave functions.
The ``interpolating functions'' introduced in Ref.\cite{GEYER} are
essentially the OFPDs which we study in this paper. 
In the early 1990s, Jain and Ralston \cite{JR} studied hard processes
involving hadron helicity flip, in terms of an ``off-diagonal transition
amplitude'' involving off-forward matrix elements of bi-quark fields in 
the nucleon. It was shown that the integral of this amplitude over the quark
four-momentum yielded elastic form factors. 
Recently, one of us \cite{JI,JIDVCS} introduced OFPDs in the study of
the spin structure of the nucleon.
The main observations in Refs.\cite{JI,JIDVCS} were that the fractions
of the spin carried by quarks and gluons can be determined from form factors
of the QCD energy-momentum tensor, and that the latter can be extracted
from the OFPDs.
Furthermore, the deeply-virtual Compton scattering (DVCS) process was
proposed \cite{JIDVCS} as a practical way to measure the new distributions.

{}From the point of view of parton physics in the infinite momentum frame,
the OFPDs have the following meaning:
if a nucleon is moving with an infinite momentum in a particular direction,
take out a parton with a certain fraction of the momentum, give it a
four-momentum transfer $\Delta^\mu$, and insert it back into the nucleon.
This is illustrated in Fig.~1.
The OFPD is then the amplitude characterizing this process.
On the other hand, from the point of view of elastic form factors, the
moments of OFPDs are form factors of twist-two quark and gluon operators.
For the spin-1 operators one has the ordinary electromagnetic and axial
form factors, while for the spin-2 operators one has the form factors of
the energy-momentum tensor.
Because the form factors of the tensor contain information about the quark
and gluon contributions to the nucleon angular momentum, the OFPDs can
provide information on the fraction of the nucleon spin carried by quark
orbital angular momentum --- a subject of considerable current interest
\cite{ORB}.

The OFPDs can be measured in diffractive processes in which the nucleon
recoils elastically after receiving some momentum transfer.
Moreover, one must have in the processes a hard, light-like momentum so
that the parton light-cone correlations are selected.
The simplest such process is DVCS \cite{JIDVCS}, in which a deeply-virtual 
photon, supplied by inelastic electron scattering, hits the nucleon and
turns into a real, high energy photon.
Such a process is easy to analyze theoretically and is similar to ordinary
deep-inelastic scattering.
More complicated processes include diffractive meson production, 
in which one must deal in addition with meson light-cone wave functions
\cite{RAD1,RAD2,HOODBHOY,BFGMS,CFS} (these have recently been shown by 
Collins et al.\cite{CFS} to be factorizable).
The best experimental facility to carry out DVCS experiments is 
the proposed ELFE \cite{ELFE}. 
However, some studies can already be made at HERA (the HERMES collaboration),
and at Jefferson Lab with a 6 GeV electron beam.

In planning future DVCS experiments, it is important to have a theoretical
estimate of the OFPDs.
The purpose of the present study is to perform a first analysis of the
OFPDs in the MIT bag model \cite{MIT}.
To be sure, the model has a number of well-known problems, including
breaking of chiral symmetry and translational invariance, absence of
explicit gluon degrees of freedom, etc.
Nonetheless, it contains quarks; it predicts well the hadron spectrum;
it gives reasonable initial input for quark distributions
\cite{STRFN,SONG,JAFFEJI};
and it can describe the electromagnetic form factors \cite{SONG,BETZ}
of the nucleon.
Other models of which we are aware have an equally long list of problems
and do not seem to provide any obvious advantage for estimating the quark
distributions.

After firstly reviewing in Section II the definitions of the off-forward
parton distributions and some of their general properties, we present in
Section III the results of the bag model for their dependence on the various
kinematic variables.
The calculation accounts for the Lorentz boost and spectator quark effects.
In Section IV we analyze the form factors of the energy-momentum tensor,
and evaluate their $t$ dependence.
Finally, conclusions are noted in Section V, and possible extensions of
this work outlined.

\section{Basics of Off-Forward Parton Distributions}

In this section we review the definitions and model independent results
for the OFPDs and their moments discussed in Ref.\cite{JI}.
Other definitions of OFPDs exist in the literature \cite{GEYER,RAD1},
however, the definition introduced in Ref.\cite{JI} has a number of
advantages, such as explicit hermiticity, and a simple connection
with local operators and their form factors.
We shall henceforth use the definition from Ref.\cite{JI}.
The OFPDs can be defined for both quarks and gluons, however, in this
paper we focus primarily on quark distributions, since the description
of the wave function of gluons in the nucleon is a much more difficult
problem.

To start, consider the bilocal operator
$\overline \psi(-\lambda n/2){\cal L} \Gamma^{\mu} \psi(-\lambda n/2)$,
where $\lambda$ is a scalar parameter, $\psi$ is a quark field of
a certain flavor, and $\Gamma^{\mu} = \gamma^{\mu}$ or
$\gamma^{\mu} \gamma_5$.
The light-like vector $n^\mu$ is proportional to $(1;0,0,-1)$, with
a coefficient depending on the choice of coordinates. 
The gauge link ${\cal L}$ is along a straight line segment extending
from one quark field to the other, which makes the bilocal operator
gauge invariant.
In the following, we work in the light-like gauge, $A\cdot n=0$,
so that the gauge link can be ignored.

One can now proceed to take the matrix element of the bilocal
operator between the nucleon states of momenta $P^\mu$ and 
$P'^\mu = P^\mu + \Delta^\mu$, where $\Delta^\mu$ is the four-momentum
transfer.
The matrix element must be expressible in terms of nucleon spinors,
Dirac matrices, and the four-vectors $P^\mu$, $\Delta^\mu$ and $n^\mu$.
Since we are only interested in the leading-twist contributions which
are proportioanl to $P^\mu$ or ${P'}^\mu$ in the infinite momentum frame,
we keep terms that are non-vanishing after multiplication by $n^\mu$
\cite{JI,JIDVCS}:
\begin{mathletters}
\label{def}
\begin{eqnarray}
\label{HEdef}
\int {d\lambda \over 2\pi} e^{i\lambda x}
	\langle P'|\overline\psi(-\lambda n/2)\gamma^\mu
			    \psi(\lambda n/2)|P \rangle
&=& H(x,\xi,t)\
	\overline u(P')\gamma^\mu u(P) \nonumber \\
&+& E(x,\xi,t)\
	\overline u(P'){i\sigma^{\mu\nu} \Delta_{\nu} \over 2M} u(P)
	+ \cdots  \ ,           \\
\label{HEtdef}
\int {d\lambda \over 2\pi} e^{i\lambda x}
        \langle P'|\overline\psi(-\lambda n/2)\gamma^\mu\gamma_5
                            \psi(\lambda n/2)|P \rangle
&=& \widetilde{H}(x,\xi,t)\
        \overline u(P')\gamma^\mu \gamma_5 u(P) \nonumber \\
&+& \widetilde{E}(x,\xi,t)\
        \overline u(P') {\Delta^\mu \gamma_5 \over 2M} u(P)
        + \cdots \ ,
\end{eqnarray}
\end{mathletters}%
where $t \equiv \Delta^2$ and $\xi \equiv -n\cdot \Delta$, with $u(P)$
the nucleon spinor, and the dots ($\cdots$) denote higher-twist
contributions.
It is possible to construct other Dirac structures that appear to be
leading-twist, however, using Gordon identities and throwing away
sub-leading terms one can always reduce these to the form in 
Eqs.(\ref{def}).
The structures in Eqs.(\ref{def}) are the same as those in the 
definition of the nucleon's elastic form factors.
Examination of the helicity structure of quark--nucleon scattering 
shows that there are exactly four independent amplitudes. 
The chiral-even distributions $H$ and $\widetilde{H}$ survive
in the forward limit in which the nucleon helicity is conserved, 
while the chiral-odd distributions $E$ and $\widetilde{E}$ arise from 
the nucleon helicity flip associated with a finite momentum transfer.

The OFPDs are depicted graphically in Fig.~1, where $k^\mu$ and $k'^\mu$
are the four-momenta of the active partons.
The physical meaning of the distributions becomes clearer if one introduces
a conjugate light-like vector $p^\mu$ of $n^\mu$, with $p\cdot n=1$.
Expanding $\overline P^\mu = (P+P')^\mu/2$ and $\Delta^{\mu}$ in terms
of the vectors $p^\mu$ and $n^\mu$ then gives:
\begin{mathletters}
\begin{eqnarray}
\overline P^\mu &=& p^\mu + (\overline M^2/2) n^\mu \ , 		\\
\Delta^\mu  &=& -\xi\ (p^\mu-(\overline M^2/2)n^\mu)
		+\Delta^\mu_\perp ,
\end{eqnarray}
\end{mathletters}%
where $\overline M^2 = \overline P^2 = M^2 - t/4$, and the spatial
components of $\overline P^\mu$ have
been chosen along the $z$-direction.
If we focus on the $p^\mu$ components of the momenta, the initial
and final nucleons have longitudinal momenta $(1+\xi/2)p^\mu$ and
$(1-\xi/2)p^\mu$, and the outgoing and incoming quarks carry
$(x+\xi/2)p^\mu$ and $(x-\xi/2)p^\mu$, respectively.
Since the nucleon cannot have negative longitudinal momentum, the
limit on $\xi$ is obviously:
\begin{equation}
	0<\xi<2 \ .
\end{equation}
A more careful analysis using $\Delta_\perp^2 > 0$ leads to the
more stringent constraint $0~<~\xi~<~\sqrt{-t}/\overline M$. 
On the other hand, since quarks cannot carry more logitudinal 
momentum than the parent nucleon, one has the constraint on $x$:
\begin{equation}
 -1 < x < 1 \ . 
\end{equation}
When $x > \xi/2$ both quark propagators in Fig.~1 represent quarks. 
When $x < -\xi/2$ they represent antiquarks.
In these regions, the OFPDs are analogous to the usual parton distributions. 
In the intermediate region, $-\xi/2 < x < \xi/2$, the quark propagators
contain one quark and one antiquark, and here the distributions resemble
a meson's wave function amplitude.

The off-forward parton distributions display characteristics of both
the forward parton distributions and nucleon form factors.
In fact, in the limit of $\Delta^\mu \rightarrow 0$, one finds \cite{JI}:
\begin{equation}
H(x,0,0) = q(x),~~~ \widetilde{H}(x,0,0) = \Delta q(x),
\end{equation}
where $q(x)$ and $\Delta q(x)$ are the forward quark and quark
helicity distributions, defined through similar light-cone correlations
\cite{JAFFEJI}.
It must be pointed out that while the $\Delta^\mu \rightarrow 0$ limit
is quite simple and natural for the OFPDs, it cannot be taken literally
for the kinematics of the DVCS process, where a finite $t$-channel
momentum transfer is essential to simultaneously maintain the initial
photon deeply-virtual, and the final state photon real.

On the other hand, the first moment of the off-forward distributions
are related to the nucleon form factors by the following sum rules
\cite{JR,JI}:
\begin{mathletters}
\label{sum1}
\begin{eqnarray}
\label{sumH}
\int^1_{-1} dx H(x,\xi,t) &=& F_1(t) \ ,          \\
\label{sumE}
\int^1_{-1} dx E(x,\xi,t) &=& F_2(t) \ ,          \\
\int^1_{-1} dx \widetilde{H}(x,\xi,t) &=& G_A(t)\ ,\\
\int^1_{-1} dx \widetilde{E}(x,\xi,t) &=& G_P(t)\ .
\end{eqnarray}
\end{mathletters}%
Here $F_1(t)$ and $F_2(t)$ are the Dirac and Pauli form factors, and
$G_A(t)$ and $G_P(t)$ are the axial-vector and pseudo-scalar form factors
of the nucleon.
The $x$-integrated distributions on the left-hand-side of Eqs.(\ref{sum1})
are in fact independent of $\xi$.
The sum rules (\ref{sum1}) provide important constraints on any model
calculation of the OFPDs.

Generalizing the sum rules (\ref{sum1}), let us multiply Eq.(\ref{HEdef})
by $x^{n-1}$ and integrate $x$ from $-1$ to $+1$, which gives:
\begin{eqnarray}
\label{nnnn}
   & &	n^{\mu_1}n^{\mu_2} \cdots n^{\mu_n}
	\langle P'|\overline \psi\ 
	i\stackrel{\leftrightarrow}{D}{}^{\mu} \cdots\ 
	i\stackrel{\leftrightarrow}{D}{}^{\mu_{n-1}}
	\gamma^{\mu_n} \psi |P\rangle			\nonumber \\
 = & &	H_n(\xi,t)\ \overline u(P') \thru n u(P)
      + E_n(\xi,t)\ \overline u(P'){i\sigma^{\mu\nu}
	n_\mu\Delta_\nu \over 2M} u(P),
\end{eqnarray}
where 
\begin{equation}
	H_n(\xi, t) = \int^1_{-1} dx\ x^{n-1}H(x,\xi,t),
\end{equation}
and likewise for $E_n(\xi,t)$.
The derivative $\stackrel{\leftrightarrow}{D}{}^{\mu}$ is defined as
\begin{eqnarray}
\stackrel{\leftrightarrow}{D}{}^{\mu}
&=& {1 \over 2} \left( \overrightarrow{D}{}^{\mu}
	       	     - \overleftarrow{D}{}^{\mu} \right),
\end{eqnarray}
where
\begin{mathletters}
\begin{eqnarray}
\overrightarrow{D}{}^{\mu}
&=& \overrightarrow\partial{}^{\mu} + i g A^{\mu},	\\
\overleftarrow{D}{}^{\mu}
&=& \overleftarrow\partial{}^{\mu} - i g A^{\mu}.
\end{eqnarray}
\end{mathletters}%
The left-hand-side of Eq.(\ref{nnnn}) is an off-forward matrix 
element of the twist-two operator
\begin{equation}
O_2^{\mu_1 \cdots \mu_2}
= \overline \psi\ 
i\stackrel{\leftrightarrow}{D}{}^{ \{ \mu_1} \cdots\ 
i\stackrel{\leftrightarrow}{D}{}^{\mu_{n-1}} \gamma^{\mu_n \}} \psi
- {\rm traces},
\end{equation}
where the braces $\{ \cdots \}$ represent symmetrization of indices.
On general grounds, a matrix element of $O_2^{\mu_1 \cdots \mu_2}$
is a sum of terms composed of form factors (which are functions of
$t$ only), appropriate Lorentz structures constructed from
$\overline P^\mu$, $\Delta^\mu$ and the Dirac matrices, and
nucleon spinors.
The $\xi$ dependence in Eq.(\ref{nnnn}) arises only from contracting
a vector $\Delta^\mu$ in any of the Lorentz structures with the null
vector $n^\mu$.
Therefore the $\xi$ dependence of the moments of the OFPDs is in the
form of polynomials.
To find the degree of the polynomials, notice that there are at most
$n$ contractions of $n^\mu$ with $\Delta^\mu$ in the $n$-th moment.
Thus $H_n(\xi,t)$ and $E_n(\xi,t)$ are polynomials of degree $n$ in $\xi$,
a result which is not obvious from the definition itself.
The same considerations apply to the bilocal operator with $\gamma_5$
dependence.

\section{A Bag Model Estimate of Off-Forward Parton Distributions}

In this section, we present a calculation of the OFPDs in a simple version
of the MIT bag model \cite{MIT}.
As mentioned in the Introduction, our choice of the MIT bag is based on
the fact that it has quark degrees of freedom, and gives reasonable results
for the electromagnetic form factors \cite{SONG,BETZ}, as well as for
polarized and unpolarized parton distributions
\cite{STRFN,SONG,JAFFEJI}.
However, being a model, it has a number of unwanted artifacts, such as
a sharp boundary, absence of gluons, and the breaking of translational
invariance and chiral symmetry.
Nonetheless, we believe that our results should provide a reasonable
first guess of the unknown distributions at a low energy scale, 
${\cal O}$(0.2 GeV$^2$).
There are, of course, many other nucleon models on the market in which
the OFPDs could be calculated, however, we see no clear reason that 
these models would be more reliable than that considered here.
The issue of evolution of the distributions to higher energy scales will
be addressed in a separate publication.

When evaluating the OFPDs and form factors, it is convenient to work in the
Breit frame, in which the initial and final momenta of the nucleon are:
\begin{eqnarray}
P_{\mu} &=& ( \overline M; -\overrightarrow \Delta/2 ),\ \ \ \ \ \ \
P'_{\mu}\ =\ ( \overline M; \overrightarrow \Delta/2 ).
\end{eqnarray}
The $t$-channel momentum transfer squared becomes:
\begin{eqnarray}
t\ =\ - \overrightarrow\Delta^2
&=& 4 \left( M^2 - \overline M^2 \right).
\end{eqnarray}
Using Eqs.(2), we then have:
\begin{eqnarray}
p^{\mu} &=& (1; 0,0,1)/(2\overline M),\ \ \ \ \ \ \
n^\mu = (1; 0, 0, -1)/\overline{M}.
\end{eqnarray}
The variable $\xi$ in this frame is therefore related to the projection
of $\overrightarrow \Delta$ in the $z$ direction,
\begin{equation}
\xi\ =\ - \Delta_z / \overline M .
\end{equation}
For calculations in a model without exact translational invariance,
the choice of frame is part of the model assumptions.
In principle, a different result could be obtained if, for instance,
one chose instead a frame where the initial nucleon was at rest.
Nevertheless, we believe that the main features of our result will be
weakly frame dependent, as in many other similar types of calculations.

Recall that the coordinate space wave function of a quark in the rest
frame of the MIT bag is given by:
\begin{eqnarray}
\label{wfn_r}
\psi(\vec r) = \sqrt{4\pi} N R^3
\left(
\begin{tabular}{c}
$j_0(\epsilon_0 r)$      \\
$i \overrightarrow \sigma \cdot\widehat{r}\ j_1(\epsilon_0 r)$
\end{tabular}
\right) \chi\ ,
\end{eqnarray}
where $R$ is the bag radius, 
$\epsilon_0 = \omega_0/R$ is the quark energy, and
$\omega_0 = 2.04$ is lowest frequency solution of the bag eigenequation,
$\tan\omega_0 = \omega_0 / (1 - \omega_0)$.
The functions $j_{0,1}$ are the spherical Bessel functions
($r \equiv |\vec r|$), and $\chi$ is the quark spinor.
The normalization $N$ is given by:
\begin{eqnarray}
N^2 &=& { \omega_0 \over 2 R^3\ (\omega_0 - 1)\ j_0^2(\omega_0) }.
\end{eqnarray}
The radius in the basic version of the bag model is given by 
the relation:\ $R M = 4 \omega_0$ \cite{MIT,JAFFEJI}.

Calculation of the OFPDs requires wave functions of a moving nucleon.
One must therefore boost the rest frame wave function (\ref{wfn_r})
to a frame moving with velocity $\vec v$. 
Including the time dependence of the quark wave function explicitly, 
$\psi(t,\vec r) = \exp(-i \epsilon_0 t)\ \psi(\vec r)$,
the effect of a Lorentz boost on the wave function can be represented
by \cite{BETZ}:
\begin{eqnarray}
\psi_{\vec v}\left( t,\vec r \right)
&=& S\left(\Lambda_{\vec v}\right)\ \
\psi\left(
	(t-\vec v \cdot \vec r) \cosh\omega;\ \ 
	\vec r + \widehat v \cdot \vec r\ (\cosh\omega-1)
		- \vec v t\ \cosh\omega \right)		\nonumber\\
&=&
\exp \left({-i \epsilon_0 (t - \vec v \cdot \vec r) \cosh\omega} \right)
							\nonumber\\
& & \times
S\left(\Lambda_{\vec v}\right)\
\psi\left(
	\vec r + \widehat v \cdot \vec r\ (\cosh\omega-1)
		- \vec v t\ \cosh\omega \right),
\end{eqnarray}
where
\begin{eqnarray}
S\left(\Lambda_{\vec v}\right)
&=& \exp\left( { \omega{ \hat v} \cdot \vec\alpha \over 2} \right)\
 =\ \cosh{\omega\over 2}\
 +\ \hat{v} \cdot \vec\alpha\ \sinh{\omega\over 2}.
\end{eqnarray}
Here the rapidity $\omega$ is related to the velocity by\
$\omega = \tanh^{-1} v$, where $v \equiv |\vec v|$.
In the Breit frame the velocity of the initial nucleon is
$\vec v~=~-\overrightarrow\Delta/2\overline{M}$, so that
\begin{eqnarray}
\cosh\omega = { \overline{M} \over M },  ~~~~~~
\sinh\omega = { |\overrightarrow\Delta| \over 2 M }.
\end{eqnarray}
Here we basically treat the independent quarks in the bag as free particles,
ignoring the fact that they are confined by the bag boundary, which again
is a part of the model assumptions.

In practical calculations, it will be more convenient to use a momentum space
wave function, $\varphi(k)$, which is simply related to the coordinate
space wave function $\psi_{\vec v}(t,\vec r)$ by a Fourier transformation:
\begin{eqnarray}
\psi_{\vec v}(t,\vec r)
&=& S\left(\Lambda_{\vec v}\right)
\int { d^3\vec{k} \over (2\pi)^3 }
\exp \left( {-i (\widetilde{\epsilon}_0 t
	     - \overrightarrow{\widetilde k} \cdot \vec r)} \right)
\varphi(\vec k),
\end{eqnarray}
where 
$\widetilde{\epsilon}_0
 = (\epsilon_0 + \vec k \cdot \vec v) \cosh\omega$,
$\overrightarrow{\widetilde k}_\perp
 = \vec k_\perp$, and
$\overrightarrow{\widetilde k}_\parallel
 = (\epsilon_0 \vec v + \vec k_\parallel) \cosh\omega$,
with 
$\vec k_\parallel = (\vec k \cdot \hat v)\ \hat v$.
The momentum space wave function is given by:
\begin{eqnarray}
\varphi(\vec k)
&=& \sqrt{4\pi} N R^3
\left(
\begin{tabular}{c}
$t_0(k)$      \\
$\overrightarrow \sigma\cdot\widehat{k}\ t_1(k)$
\end{tabular}
\right) \chi\ ,
\end{eqnarray}
where $k \equiv |\vec k|$,
and the functions $t_{0,1}$ are given by:
\begin{mathletters}
\begin{eqnarray}
t_0(k)
&=& { j_0(\omega_0)\ \cos(k R)
    - j_0(k R)\ \cos\omega_0
    \over \omega_0^2 - \vec k^2 R^2 }\ ,	\\
t_1(k)
&=& { j_0(k R)\ j_1(\omega_0)\ k R
    - j_0(\omega_0)\ j_1(k R)\ \omega_0
    \over \omega_0^2 - \vec k^2 R^2 }\ .
\end{eqnarray}
\end{mathletters}%

One of the most important issues in calculating the off-forward matrix 
elements of single-particle operators in independent particle models like
the bag is momentum conservation. 
In the following, we discuss this issue in some detail, which in the end
will motivate the approach we take in performing the calculation. 
In independent particle models, strictly speaking the form factors 
of any one-body operator must be zero. 
Since the momentum transfer through the one-body operator affects the
active quark only, the remaining spectator quarks maintain their original 
states.
On the other hand, if a momentum transfer $\vec\Delta$ is given to the
whole nucleon, each of the quarks must receive a momentum transfer
$\vec\Delta/3$, since before and after the interaction, the model
nucleon must move as a whole. 
Thus, due to the momentum mismatch, the form factors must vanish. 
In the realistic situation, however, the nucleon wave function contains 
correlations.
The momentum transfer injected to a single quark is in turn transferred
through correlations to the other constituents, and asymptotically is
equally shared among them.
In independent particle models, however, these vital correlations
needed for form factor calculations are missing.

In the literature, several common approaches have been adopted to 
deal with this issue. 
In one approach, model wave functions with no definite 
center-of-mass momentum are used \cite{BETZ,SONG}.
Form factors are calculated 
from a Fourier component of the single particle operator, 
$\widehat O(\vec{\Delta}) = \int d^3\vec{r}\ e^{i\vec{r}\cdot \vec{\Delta}}
\widehat O(\vec{r})$.
In this type of calculation, the momentum 
transfer to the model nucleon and to the individual quarks
is not well defined. If a boosted single-particle 
wave function is used, roughly speaking, the momentum transfer
to each spectator quark is $\vec{\Delta}/3$. On the other hand,
the momentum transfer through the active quark is approximately
$(1-\epsilon_0/M)\vec{\Delta}$, which arises from the combined
effects of boost and the action of the single-particle 
operator. 
(The relative sign of the two effects appears somewhat counterintuitive, 
but is nonetheless correct.)

In principle, a better approach would be to use initial and
final nucleon wave functions with definite momentum, which 
can be approximately obtained, for instance, through the 
Peierls-Yoccoz projection \cite{PY} or the center-of-mass 
freedom separation method\cite{WSY}. 
In such calculations, all quarks share equally the  
momentum transfer to the nucleon.
In particular, the active quark in which the single-particle
operator acts is injected with a momentum $\vec{\Delta}/3$ 
only. Since the state of the two spectator quarks also 
changes from the initial to the final nucleon, the effective 
operator which induces such a transition is actually 
a three-body operator.

A calculation of the OFPDs incorporating the effects of Lorentz boosts
and projections is rather involved, and in practice not particularly
illuminating.
Instead, we consider a simpler alternative, by modifying 
the momentum transfer through the active quark in the approach 
of Ref.\cite{BETZ}. Namely, we let the effetive 
momentum transfer through the active quark be $\eta \vec\Delta$, 
where $\eta$ is taken to be a parameter. As mentioned
above, $\eta = 1-\epsilon_0/M$ in an unprojected calculation.
Ultimately, $\eta$ in our calculation will be determined by 
fitting the electromagnetic form factors of the nucleon, but 
one can expect $\eta$ to be around 1/3, in the spirit of the
momentum-projected calculation discussed above.

The matrix element of the bilocal operator is calculated using the
boosted wave function of the active quark,
\begin{eqnarray}
\label{singleq}
& &
2 \overline{M}
\int{d\lambda \over 2\pi} e^{i \lambda x}
\int d^3r\ e^{i \vec\Delta \cdot \vec r}\ \ 
\overline{\psi}_{-\vec v}(-\lambda n/2 + \vec r)\ \Gamma\
\psi_{\vec v}(\lambda n/2 + \vec r)			\nonumber\\
&=&
{2 \overline{M} \over \cosh\omega}
\int{d^3k \over (2\pi)^3}\
\delta \left( x - n_- (\tilde{k}_+ + {\Delta_+ \over 2}) \right)
\left\{
\varphi^{\dagger}(k')\ S(\Lambda_{-\vec v})\ \gamma_0 \Gamma\
S(\Lambda_{\vec v})\ \varphi(k)
\right\},
\end{eqnarray}
where $\Gamma = \thru n$ or $\thru n \gamma_5$,
and
$k' \equiv |\vec k'|$,
$\vec k' = \vec k + \overrightarrow{\widetilde{\Delta}}$.
The effective momentum transfer $\overrightarrow{\widetilde{\Delta}}$ 
is given by:
\begin{eqnarray}
\overrightarrow{\widetilde\Delta}
&=& \eta\ { \overrightarrow{\Delta} \over \cosh\omega }.
\end{eqnarray}
Choosing for simplicity $\Delta_y = 0$, and using cylindrical coordinates
to perform the $k$ integration, the $\delta$-function in Eq.(\ref{singleq})
reduces to a constraint on the $z$-component of $\vec k$:
\begin{eqnarray}
k_z &=&
{ \overline M \over 1 - (\cosh\omega-1) \Delta_z^2/t }		\nonumber\\
&\times&
\left[ x - {1 \over 2 \overline M}
   \left( (2 \epsilon_0 + \widetilde\Delta_z) \cosh\omega
	+ |\overrightarrow{\widetilde\Delta}| \sinh\omega
	- {2 \Delta_x \Delta_z \over t } k_x (\cosh\omega-1)
   \right)
\right] .
\end{eqnarray}
Evaluating the expression in the braces in Eq.(\ref{singleq}) explicitly
for $\Gamma = \thru n$, and equating the spin independent components on
both sides, leads to:
\begin{eqnarray}
\label{HE1}
& & H(x,\xi,t) + { t \over 4 M^2 } E(x,\xi,t)		\nonumber\\
&=& Z^2(t) \left(4 \pi N^2 R^6\right)
{ \overline{M} \over 1 - (\cosh\omega-1) \Delta_z^2/t }
\int { dk_\perp\ d\varphi \over (2\pi)^3 }\ k_\perp		\nonumber\\
&\times&
\left\{ 
t_0(k)\ t_0(k')
+\
\left[
k'_z \cosh\omega
+ {2 \Delta_z \over t} \vec k' \cdot \overrightarrow\Delta\ 
  \sinh^2{\omega\over 2}
\right]
{ t_0(k)\ t_1(k') \over k' }
\right.					\nonumber\\
& & +\
\left[
k_z\ \cosh\omega
+ {2 \Delta_z \over t}\
  \vec k \cdot \overrightarrow\Delta\ 
  \sinh^2{\omega\over 2}
\right]
{ t_1(k)\ t_0(k') \over k }				\nonumber\\
& & \left. +\
\left[
\vec k \cdot \vec k'
- \widehat\Delta \cdot \left( \vec k\ \widetilde{\Delta}_z 
	  		- \overrightarrow{\widetilde\Delta}\ k_z 
		       \right) \sinh\omega
\right]
{ t_1(k)\ t_1(k') 
  \over k k' }
\right\},
\end{eqnarray}
where the effects of the spectator quarks are included in the
factor $Z(t)$ \cite{BETZ}:
\begin{eqnarray}
Z(t) &=& {N^2 \over \cosh\omega}
\int_0^R dr\ r^2\
j_0 \left( \epsilon_0 |\vec\Delta| r / \overline{M} \right)
	\left( j_0^2(\epsilon_0 r) + j^2_1(\epsilon_0 r) \right).
\end{eqnarray}
If one compares the $\sigma_y$ components in Eq.(\ref{singleq}),
on the other hand, a different combination of $H$ and $E$ arises:
\begin{eqnarray}
\label{HE2}
& & H(x,\xi,t) + E(x,\xi,t)				\nonumber\\
&=& Z^2(t)\ \left( 4 \pi N^2 R^6 \right)\
{ 2 M \overline{M} \over 1 - (\cosh\omega-1) \Delta_z^2/t }
\int { dk_\perp\ d\varphi \over (2\pi)^3 } k_\perp	\nonumber\\
&\times&
\left\{
{ \sinh\omega \over |\overrightarrow\Delta| }
t_0(k)\ t_0(k')
+\
\left[
{ k'_x \over \Delta_x } \cosh\omega
- {2 \Delta_z \over t}
  { ( \vec k \times \vec\Delta )_y \over \Delta_x }\
  \sinh^2{\omega\over 2}
\right]
{ t_0(k)\ t_1(k') \over k' }
\right.							\nonumber\\
& & -\
\left[
{ k_x \over \Delta_x } \cosh\omega
- {2 \Delta_z \over t}
  { ( \vec k \times \vec\Delta )_y \over \Delta_x }\
  \sinh^2{\omega\over 2}
\right]
{ t_1(k)\ t_0(k')
  \over k }						\nonumber\\
& & \left. +\
\left[
{ ( \vec k \times \vec{\widetilde\Delta} )_y
  \over \Delta_x }
- { \vec k \cdot \vec k' - 2 k_y^2 \over |\vec\Delta| }
  \sinh\omega
\right]
{ t_1(k)\ t_1(k')
  \over k k' }
\right\}.
\end{eqnarray}
Expressions for the individual functions $H$ and $E$ are then obtained
by solving Eqs.(\ref{HE1}) and (\ref{HE2}) directly.

For the helicity dependent case $\Gamma =  \thru n \gamma_5$
again one can obtain two independent combinations of $\widetilde{H}$
and $\widetilde{E}$ by comparing different spin components in 
Eq.(\ref{singleq}).
Equating coefficients of $\sigma_z$ leads to:
\begin{eqnarray}
\label{HE3}
& & 2 \left(1 - {\Delta_z^2 \over 4 \overline{M} (M+\overline{M})}\right)
	\widetilde{H}(x,\xi,t)
- { \Delta_z^2 \over 2 M \overline M }
	\widetilde{E}(x,\xi,t)				\nonumber\\
&=& Z^2(t)\ \left( 4 \pi N^2 R^6 \right)\
{ 2 M \over 1 - (\cosh\omega-1) \Delta_z^2/t }
\int { dk_\perp\ d\varphi \over (2\pi)^3 } k_\perp	\nonumber\\
&\times&
\left\{
\left[
\cosh\omega + {2 \Delta_z^2 \over t} \sinh^2{\omega\over 2}
\right]
t_0(k)\ t_0(k')
+\
\left[
k'_z 
+ { \Delta_x\ k'_x \over |\vec\Delta| } \sinh\omega
\right]
{ t_0(k)\ t_1(k') \over k' }
\right.					\nonumber\\
& & +\
\left[
k_z - { \Delta_x k_x \over |\vec\Delta| } \sinh\omega
\right]
{ t_1(k)\ t_0(k') \over k }				\nonumber\\
& & +\
\left[
\left( 2 k_z k'_z - \vec k \cdot \vec k' \right) \cosh\omega
\right.							\nonumber\\
& & \hspace*{0.5cm} \left.\left.
+ {2 \Delta_z \over t}
  \left( k_z (\vec k + \vec k') \cdot \vec\Delta 
       - \Delta_z \vec k^2
  \right)
  \sinh^2{\omega\over 2}
\right]
{ t_1(k)\ t_1(k') \over k k' }
\right\},
\end{eqnarray}
while the $\sigma_x$ components give:
\begin{eqnarray}
\label{HE4}
& &
- { \Delta_x \Delta_z \over 2 \overline{M} (M+\overline{M}) }
	\widetilde{H}(x,\xi,t)
- { \Delta_x \Delta_z \over 2 M \overline M }
	\widetilde{E}(x,\xi,t)				\nonumber\\
&=& Z^2(t)\ \left( 4 \pi N^2 R^6 \right)\
{ 2 M \over 1 - (\cosh\omega-1) \Delta_z^2/t }
\int { dk_\perp\ d\varphi \over (2\pi)^3 } k_\perp	\nonumber\\
&\times&
\left\{
\left[ {2 \Delta_x \Delta_z \over t}
	\sinh^2{\omega \over 2}
\right]
t_0(k)\ t_0(k')
+\
\left[
k'_x - { \Delta_x\ k'_z \over |\vec\Delta| } \sinh\omega
\right]
{ t_0(k)\ t_1(k') \over k' }
\right.					\nonumber\\
& & +\
\left[
k_x + { \Delta_x\ k_z \over |\vec\Delta| } \sinh\omega
\right]
{ t_1(k)\ t_0(k') \over k }				\nonumber\\
& & +\
\Big[
\left( k_x k'_z + k_z k'_x \right) \cosh\omega		\nonumber\\
& & \ \ \ \ \left.
+ {2 \Delta_z \over t}
  \left( k_x (\vec k + \vec k') \cdot \vec\Delta
       - \Delta_x \vec k^2
  \right)
  \sinh^2{\omega\over 2}
\Big]
{ t_1(k)\ t_1(k') \over k k' }
\right\}.
\end{eqnarray}

By fixing the bag radius to be $R = 4 \omega_0/M$ \cite{MIT,JAFFEJI},
the model then has essentially one parameter, $\eta$.
This can be constrained by comparing the model predictions for the
Sachs electric and magnetic form factors of the proton,
\begin{mathletters}
\begin{eqnarray}
G_E(t) &=& F_1(t) + { t \over 4 M^2 } F_2(t),	\\
G_M(t) &=& F_1(t) + F_2(t),
\end{eqnarray}
\end{mathletters}%
with the available data \cite{DATA1,DATA2}.
The form factors can be obtained from Eqs.(\ref{sumH}) and (\ref{sumE}) 
by integrating the $H$ and $E$ distributions in Eqs.(\ref{HE1}) and
(\ref{HE2}) directly.
Note that the results are in fact equivalent to those given for $G_E$
and $G_M$ in coordinate space in Ref.\cite{BETZ}.
Furthermore, in a calculation for the proton with an SU(6) symmetric wave 
function, the combinations of the electric, baryonic, and axial charge 
factors appear in such a way as to reduce the final result to that for
a single quark.
In Fig.~2 we show the predicted $t$ dependence of $G_E$ for two values
of the parameter $\eta$ ($\eta=0.35$ and 0.55).
The $t$ dependence of the magnetic form factor $G_M$ is shown in Fig.~3.
In both cases the bag model results are in quite good agreement with
the data.
The small-$|t|$ data (including the charge radius and magnetic moment)
do favor the larger value $\eta=0.55$, while a better fit can be achieved
at large $|t|$ with $\eta=0.35$.
Note that while $G_E(0)$ is independent of $R$ and $\eta$, the value
of the magnetic moment $G_M(0)$ depends on the bag radius and $\eta$.
This shows that different prescriptions of calculating form factors
within the model can give different answers even at small momentum 
transfers.
This is just one of the artifacts of the explicit breaking of 
translational invariance in the bag model.

Having fixed the model parameters, one can now calculate the individual
OFPDs as a function of $x$ and $\xi$, for different values of $t$.
Again, assuming the SU(6) wave function for the proton, one multiplies
the right-hand-side of Eq. (\ref{HE1}) by a factor 2 (1) and that of
Eqs. (\ref{HE2}), (\ref{HE3}) and (\ref{HE4}) by a factor 4/3 (--1/3)
to solve for the up (down) quark distributions.
In Figs.~4(a) and (b) we first show the distributions at $t=0$ 
(and $\xi=0$) for $-1 < x < 1$, for both the $u$ and $d$ quark flavors.
Because the small-$|t|$ form factors are better described with a larger 
$\eta$ value, for consistency we use here $\eta=0.55$.
Note that the distributions $H$ and $\widetilde H$ in Fig.~4(a) for 
$u$ and $d$ quarks are just the forward unpolarized $u(x)$ and $d(x)$,
and polarized $\Delta u(x)$ and $\Delta d(x)$ parton distributions,
respectively.
Both $H$ and $\widetilde H$ are in fact independent of $\eta$.
Furthermore, the first moments of $H$ and $\widetilde{H}$ at $t=0$
explicitly satisfy the normalization conditions (equal to 2 and 1
for unpolarized $u$ and $d$ quarks, and $4/3\ g_A$ and $-1/3\ g_A$,
where $g_A = 0.65$ for a quark, in the polarized case, respectively).
The tensor and pseudoscalar distributions $E$ and $\widetilde E$,
however, shown in Fig.~4(b), do depend rather strongly on $\eta$.
The peak value of $\widetilde E$ for the $u$ quark for instance would
be $\sim 6$ for $\eta = 1-\epsilon_0/M = 0.75$ (with $R=4\omega_0/M$).
The first moment of $E$ at $t=0$, summed over both the $u$ and $d$
flavors, is equal to $G_M(0)-1$ (as in Fig.~3), while the first moment
of the flavor singlet $\widetilde E$ is $G_P(0) \approx 6$ for 
$\eta = 0.55$.
Note also that the calculated $x$ distributions do not vanish entirely
at $x=1$, which simply reflects the fact that the initial and final
nucleons are not good momentum eigenstates.
However, the effect is only slightly noticeable for $\widetilde E$,
and is negligible for the other distributions.

In Figs.~5--10 the distributions for both the $u$ and $d$ quarks are
shown as a function of $x$ and $\xi$, for $t=-1$ and $t=-2$ GeV$^2$.
Throughout we use the value $\eta=0.35$ on account of the better
agreement with the form factor data at large $|t|$ (see Figs.~2--3).
One can see that the dependence on $\xi$ is quite weak.
According to the discussion in the last section, this means that the
form factors of the twist-two operators associated with the structure\
$\overline P^{\mu_1} \cdots \overline P^{\mu_{n-1}}
\overline u(P')\gamma^{\mu_n} u(P)$\
dominate over other form factors.
We do not have a simple explanation for this. 
The $t$ dependence of OFPDs, however, is rather strong, as expected
from the above form factor behavior.
Experimentally, the most interesting region for DVCS is $1<-t<2$ GeV$^2$.
For too small $|t|$, QED radiative effects mask the processes sensitive
to the OFPDs.
On the other hand, for too large $|t|$, the distributions become too
small to be measurable.

At asymptotically large $|t|$, the OFPDs can be analyzed using
perturbative QCD.
Again, it is the leading-twist light-cone wave function which
determines the behavior of the helicity conserving distributions.
We expect therefore $H(x,\xi,t)$ and $\widetilde H(x,\xi,t)$
to fall like $1/t^2$ as $-t \rightarrow \infty$.

\section{Form Factors of the Energy-Momentum Tensor}

In this section we review the role of the form factors of the QCD
energy-momentum tensor played in the spin structure of the nucleon,
and their relation with the OFPDs.
In particular, we present the first model calculation of the form factors
of the quark part of the energy-momentum tensor.

Since the publication of the EMC measurement \cite{EMC} of the fraction
of the proton's spin carried by quarks, the spin structure of the nucleon
has been studied extensively in the literature.
A deeper understanding of the problem, however, requires one to examine
more closely the angular momentum operator in QCD.
This can be written as a sum of quark and gluon contributions \cite{JI}:
\begin{equation}
\overrightarrow J_{\rm QCD}
= \overrightarrow J_{q} + \overrightarrow J_g \ ,
\end{equation}
where
\begin{mathletters}
\begin{eqnarray}
\overrightarrow J_q
&=& \int d^3r ~\vec r \times \overrightarrow T_q		\nonumber\\
 &=& \int d^3r ~\left[
    {1 \over 2} \psi^\dagger \overrightarrow \Sigma \psi
    + \psi^\dagger ~\vec r \times (-i \overrightarrow D)\psi
                \right] \ ,					\\
\overrightarrow J_g
&=& \int d^3r ~\vec r \times
        (\overrightarrow E \times \overrightarrow B) \ .
\end{eqnarray}
\end{mathletters}%
where summations over flavor and color indices are implicit, and
$\overrightarrow T_q$ and $\overrightarrow E \times \overrightarrow B$
are the quark and gluon momentum densities, respectively.
The Dirac spin-matrix is denoted by $\overrightarrow \Sigma$, and
$\overrightarrow D$ is the covariant derivative.
By analogy with the magnetic moment, the separate quark and gluon 
contributions to the nucleon spin can be obtained from the form factors
of the momentum density, or equivalently the QCD energy-momentum tensor
at zero momentum transfer.

The symmetric, conserved, gauge-invariant energy-momentum tensor
$T^{\mu\nu}$ of QCD can be separated into quark and gluon components:
\begin{equation}
T^{\mu\nu} = T^{\mu\nu}_q + T^{\mu\nu}_g \ ,
\end{equation}
where the quark part is:
\begin{equation}
\label{Tmunu_q}
T^{\mu\nu}_q
= {1 \over 2}
\left( \overline \psi \gamma^{\{\mu} i \uprightarrow{ D^{\nu\}}} \psi 
     + \overline \psi \gamma^{\{\mu} i \upleftarrow{ D^{\nu\}}}  \psi
\right) \ ,
\end{equation}
and the gluon part is:
\begin{equation}
T^{\mu\nu}_g
= {1\over 4}g^{\mu\nu} F^2 - F^{\mu\alpha}F^\nu_{~\alpha} \ .
\end{equation}
Using Lorentz covariance and invariance under the discrete symmetries,
one can expand the matrix elements of $T^{\mu\nu}_{q,g}$ in terms of
four form factors:
\begin{eqnarray}
\label{Tmunu_me}
\langle P' | T^{\mu\nu}_{q,g} | P \rangle
&=& \overline u(P')
\Big[ A_{q,g}(t)\ \gamma^{\{\mu} \overline P^{\nu\}}\
   +\ B_{q,g}(t)\ \overline P^{\{\mu}
        i\sigma^{\nu\}\alpha}\Delta_\alpha/2M           \nonumber \\
& & \hspace*{1cm}
   +\ C_{q,g}(t)\ (\Delta^\mu \Delta^\nu - g^{\mu\nu}t)/M\
   +\ \overline C_{q,g}(t)\ g^{\mu\nu}M\Big] u(P) \ ,
\end{eqnarray}
where the braces $\{ \cdots \}$ on the superscripts denote
symmetrization.
Substituting the above into the nucleon matrix element of
$\overrightarrow J_{q,g}$, one finds fractions of the nucleon
spin carried by quarks, $J_q$, and gluons, $J_g$ \cite{JI},
\begin{mathletters}
\begin{eqnarray}
\label{JAB}
J_{q,g} &=& {1\over 2} \Big( A_{q,g}(0)+B_{q,g}(0) \Big)\ ,     \\
\label{JqJg}
J_q + J_g &=& {1 \over 2}\ ,
\end{eqnarray}
\end{mathletters}%
where in the rest frame of the nucleon
\begin{eqnarray}
J_{q,g} \vec S &=& \langle P | \overrightarrow J_{q,g} | P \rangle, 
\end{eqnarray}
with $\vec{S}$ the polarization vector of the nucleon.
This role of the form factors of the total energy-momentum tensor has
been first noted by Jaffe and Manohar \cite{JM}.

Measuring the form factors of the energy-momentum tensor in practice
is very difficult, partly because there is no fundamental probe which
couples to them.
(The graviton does, but only to the total tensor.)
Because of asymptotic freedom, the form factors can be measured through
deep-inelastic sum rules, as explained in Ref.\cite{JI}.
According to our definition of OFPDs, it is simple to show that:
\begin{eqnarray}
\label{sumxH}
\int_{-1}^1 dx\ x H(x,\xi,t)
&=& A(t) + \xi^2\ C(t)\ ,	\\
\label{sumxE}
\int_{-1}^1 dx\ x E(x,\xi,t)
&=& B(t) - \xi^2\ C(t)\ .
\end{eqnarray}
Combining these one obtains:
\begin{eqnarray}
\label{sum2}
\int_{-1}^1 dx\ x \Big( H(x,\xi,t) + E(x,\xi,t) \Big)
&=& A(t) + B(t)\ ,
\end{eqnarray}
where the $\xi$ dependence drops out!

For the spin structure of the nucleon only small values of $t$ are relevant.
However, reaching small $t$ in real experiments will be difficult; so
some knowledge of the $t$ dependence of the form factors is essential.
In the remainder of this section, we will present a detailed calculation
of the form factors of the (quark flavor-singlet part of the) 
energy-momentum tensor in the bag model. 
In principle, one can already obtain the form factors from the sum rules
(\ref{sumxH})--(\ref{sum2}).
However, here we will present an independent derivation, from which not only
can one check the calculations on OFPDs, but also obtain the form factors
not accessible from the above sum rules.

The bag energy-momentum tensor is a sum of quark and empty-bag contributions.
The quark part can be written:
\begin{equation}
T^{\mu\nu}_{q, Bag}
= {1 \over 2}
\left( \overline \psi \gamma^{\{\mu} i \uprightarrow{ \partial^{\nu\}}}
\psi  + \overline \psi \gamma^{\{\mu} i \upleftarrow{ \partial^{\nu\}}}
\psi\right) \ ,
\end{equation}
where $\psi$ is the quark field in the bag.
Using the bag's equation of motion, $i \thru \partial \psi = \delta(r-R)$,
it is easy to see that $T^{\mu\nu}_{q, Bag}$ is traceless.
{}From the general theorem proved in Ref.\cite{MASS}, the bag quarks
contribute 3/4 of the nucleon's mass.
However, as we shall see below, the quarks in the bag carry all of the
momentum of the nucleon.

The form factors $A, B, C$ and $\overline C$ are calculated from
the matrix elements of $T^{\mu\nu}_{q, Bag}$ by taking various
components of $T^{\mu\nu}_{q, Bag}$ and comparing with the form 
in Eq.(\ref{Tmunu_me}).
Taking $\overrightarrow\Delta$ to be in the $z$-direction, one finds
from the $T^{0i}_{q, Bag}$ components:
\begin{eqnarray}
\label{AB}
A(t) + B(t)
&=& 2\ Z^2(t)\ \left( 4 \pi N^2 R^6 \right)\
\int{d^3k \over (2\pi)^3}
\left\{
{ k_x^2 \over \cosh\omega } { t_1(k) t_1(k') \over k k' }
\right.                                                 \nonumber\\
& & 
+\ { M \over |\vec\Delta| }
\left(
{ \epsilon_0 \over M } - { \eta\ t \over 4 \overline M^2 }
\right)
\left[
\left(
t_0(k) t_0(k') - { k_z k'_z \over k k' } t_1(k) t_1(k') 
\right) \sinh\omega\
\right.                                                 \nonumber\\
& & \hspace*{3cm}
\left.
\left.
+\
\left(
{ k'_z \over k' } t_0(k) t_1(k') - { k_z \over k } t_1(k) t_0(k')
\right) \cosh\omega
\right]
\right\}.
\end{eqnarray}
In the case of no Lorentz boost, $\omega = 0$, and the combination
$A+B$ is unity at $t=0$.
According to Eq.(\ref{JAB}), this represents the fact that the total
angular momentum (spin plus orbital) of quarks in the bag is 1/2.
In Fig.~11 this is illustrated by the dotted curve.
With a Lorentz boost and a momentum transfer fraction to the quark
$\eta < 1 - \epsilon_0/M$, the value of $A(0)+B(0)$ is no longer unity.
This can be understood from the fact that the boosted bag wave function
does not have the correct Lorentz symmetry.
For a smaller value $\eta = 0.35$ (solid) one finds that $A+B$ at $t=0$
is now $\approx 0.5$.
However, as discussed in the previous section, the small-$t$ form factors
are more accurately described with a larger $\eta$, $\eta=0.55$ (dashed)
which gives $A(0)+B(0) \approx 0.7$.

In dispersion theory, the $t$ dependence of form factors is controlled
by the nearest singularities in $t$-channel.
For $A(t)+B(t)$, the quantum number of the channel is the exotic state
with $J^{PC} = 1^{-+}$.
There is no conclusive evidence at the present time for the existence 
of resonances in this channel, although theoretical investigations
indicate that hybrid $q\bar qg$ mesons could exist in the mass range
of 1.3 to 1.9 GeV \cite{BALITSKY}.
If the dispersive behavior of $A(t)+B(t)$ is dominated by large-mass
resonances, then $A+B$ will vary slowly with $t$, at least noticeably
slower than the electromagnetic form factor.
Although the bag calculation does indicate such a trend, the evidence
is not strong.
We suspect therefore that either the $t$ dependence of the bag prediction
is not reliable, or multi-pion cuts in the form factor are important. 
Further study in this direction is called for.

To obtain the individual form factors $A$, $B$, $C$, $\overline C$,
one can take other components of the tensor $T^{\mu\nu}_{q, Bag}$.
The $T^{00}_{q, Bag}$ component gives:
\begin{eqnarray}
\label{ABCCb}
& & A(t) + {t \over 4 M^2} B(t)
- {t \over M^2} C(t) + \overline{C}(t)		\nonumber\\
&=& 3\ Z^2(t)\ \left( 4 \pi N^2 R^6 \right)\
\left( { \epsilon_0 \over M }
     - { \eta\ t \over 4 \overline{M}^2 }
\right)
\int{d^3k \over (2\pi)^3}
\left\{
t_0(k) t_0(k')
+ \widehat k \cdot \widehat k'\ t_1(k) t_1(k')
\right\},
\end{eqnarray}
where the factor of 3 is the number of valence quarks,
and from the $T^{xx}_{q, Bag}$ (or $T^{yy}_{q, Bag}$) components one gets:
\begin{eqnarray}
\label{CCb}
{t \over M^2} C(t) - \overline{C}(t)
&=& 3\ Z^2(t)\ \left( 4 \pi N^2 R^6 \right)\
{ 1 \over M \cosh\omega }
\int{ d^3k \over (2\pi)^3 } k_x^2               \nonumber\\
& & \times
\left\{
\left[
{ t_0(k) t_1(k') \over k' } + { t_1(k) t_0(k') \over k }
\right] \cosh\omega\
+\ 
{ t_1(k) t_1(k') \over k k' } |\overrightarrow{\widetilde\Delta}|
\sinh\omega
\right\}.
\end{eqnarray}
Finally, the $T^{zz}_{q, Bag}$ components give a fourth equation:
\begin{eqnarray}
\label{Cb}
\overline{C}(t)
&=& -3\ Z^2(t)\ \left( 4\pi N^2 R^6 \right)\ {1 \over 2 M}
\int{d^3k \over (2\pi)^3}
(k_z + k'_z)
\left\{
{ k'_z \over k' } t_0(k) t_1(k')\
+\ 
{ k_z \over k } t_1(k) t_0(k')
\right\}.
\end{eqnarray}
Solving Eqs.(\ref{AB})--(\ref{Cb}), we plot in Fig.~12 the resulting
form factors as a function of $t$ for $\eta=0.35$. 
All form factors fall off monotonically as $-t$ increases, with $A(t)$
being the largest (and positive), and others relatively smaller
(and negative). 
Note that for a radius $R = 4\omega_0/M$, the form factor $A(0) = 1$,
which according to the definition of the form factors simply reflects
the fact that all the momentum of the nucleon is carried by quarks.

The trace condition on the quark part of the bag energy-momentum 
tensor implies that not all of the form factors are independent. 
In fact, it is easy to show that,
\begin{eqnarray}
A(t) + { t \over 4 M^2 } B(t) 
- { 3 t \over M^2 } C(t) + 4 \overline{C}(t)
&=& 0\ .
\end{eqnarray}
We have checked that this relation is in fact satisfied to a very good
approximation in the bag model, despite the violation of Lorentz symmetry
after boosts.
The above equation indicates that the normalization of the non-conserving
form factor is fixed to $\overline C(0) = 1/4$.
It must be pointed out, however, that the above relation is not true in
QCD because the renormalized quark part of the energy-momentum tensor
has a trace anomaly \cite{CDJ,NIELSON}.

\section{Conclusion}

In this paper we have presented a detailed study of a new class of 
nucleon observables, the off-forward parton distributions.
The physical significance of the distriubtions has been explained in a
partonic language, and their relations to form factors of twist-two
operators made explicit.
The $\xi$ dependence of the moments of the OFPDs is found to have a rather
simple polynomial form.
We have made the first model calculation of the distributions using the
MIT bag model, with the specific details of the model fixed by requiring
one to reproduce the electromagnetic form factors of the proton, and the
gross features of the parton distributions at a low energy scale.
In relation to the spin structure of the nucleon, we have also studied
the form factors of the energy-momentum tensor in the bag model, focusing
on the $t$ dependence in the range of 0 to --2 GeV$^2$.

The model calculation includes the effects of Lorentz boosts of the
quark wave function, and spectator quarks.
The $\xi$ dependence of the distributions turns out to be extremely weak,
indicating that the form factors of the twist-two operators associated
with the Lorentz structure 
$\overline{P}^{\mu_1} \cdots \overline{P}^{\mu_{n-1}} 
\overline u(P') \gamma^{\mu_n} u(P)$ are dominant.
The $t$ dependence of the OFPDs, on the other hand, is much stronger, 
and roughly follows that of the elastic form factors.
Our results for the combination $A+B$ of the tensor form factors provide
the first concrete indication of their possible $t$ dependence, which
will be important for the extraction of the value at $t=0$, since this
gives the total angular momentum carried by quarks in the nucleon.

An important issue to be addressed in future is the $Q^2$ evolution
of the OFPDs.
It is well known that the bag model calculations of the forward parton
distributions are valid only at very low energy scales, 
${\cal O}(0.2)$ GeV$^2$, therefore the OFPDs calculated in this 
paper cannot be used directly to estimate the experimental cross
sections.
The evolution equations for the off-forward distributions, which
interpolate the Brodsky-Lepage \cite{BL} and DGLAP \cite{DGLAP}
evolution equations, have been derived in Refs.\cite{JIDVCS,RAD1,RAD2}.
The technology for solving these new evolution equations is currently
being developed.
Apart from the evolution, the model calculations of the OFPDs could
also be refined in future by including Lorentz boosts together with
momentum projections.

The most relevant experiment from which the off-forward distributions
and tensor form factors can be extracted is deeply-virtual Compton
scattering \cite{JI,JIDVCS}.
In practice, the DVCS process will be overwhelmed at small $t$ by the
Bethe-Heitler process, whose cross section has a QED infrared divergence
at $t=0$, so that one cannot go to too small $t$.
One may get around this to some extent by subtracting the (calculable)
Bethe-Heitler contribution, or by isolating the virtual Compton
and Bethe-Heitler interference term \cite{VCS} through single-spin
asymmetry or combined lepton-antilepton scattering \cite{JIDVCS}.
Other processes which can also provide information on the OFPDs include
diffractive $\rho$ and $J/\psi$ production \cite{RAD2,HOODBHOY,BFGMS,CFS},
which are sensitive to the off-forward gluon distribution.
A series of dedicated experiments at suitable machines, such as HERMES,
ELFE, or at Jefferson Lab, can in future explore these new distributions,
and thus provide valuable information on the distribution of spin in the
nucleon.

\acknowledgements

This work was supported by the DOE grant DE-FG02-93ER-40762.
X. Song was supported in part by the DOE grant
DE-FG02-96ER-40950, Institute of Nuclear and Particle Physics,
Department of Physics, University of Virginia, and the Commonwealth
of Virginia.


\begin{figure}
\label{fig1}
\epsfig{figure=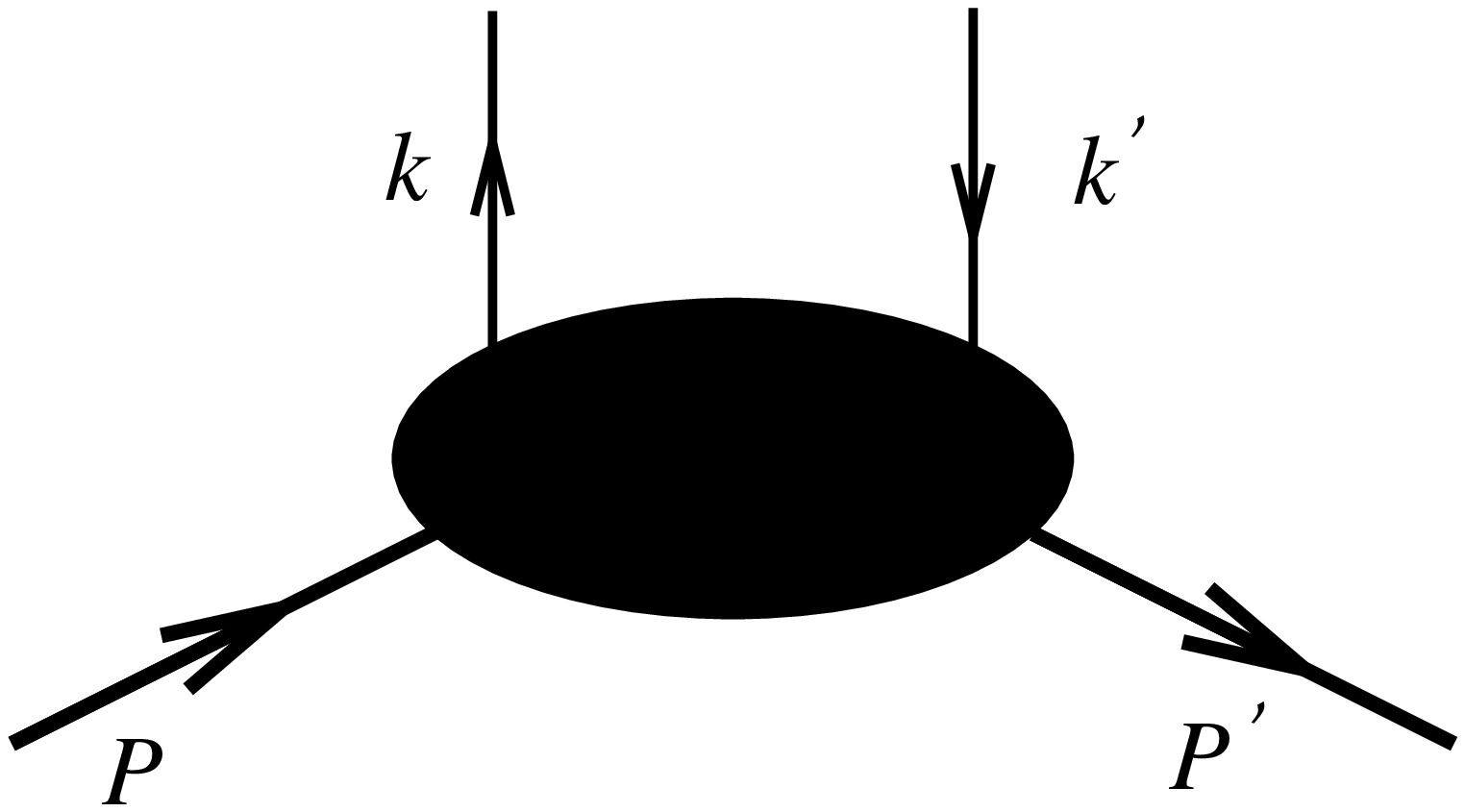,height=8cm}
\caption{Off-forward parton distribution.
	$P$ and $P'$ are the initial and final state nucleon momenta,
	and $k$ and $k'$ are the active quark momenta.}
\end{figure}

\newpage

\begin{figure}
\label{fig2}
\epsfig{figure=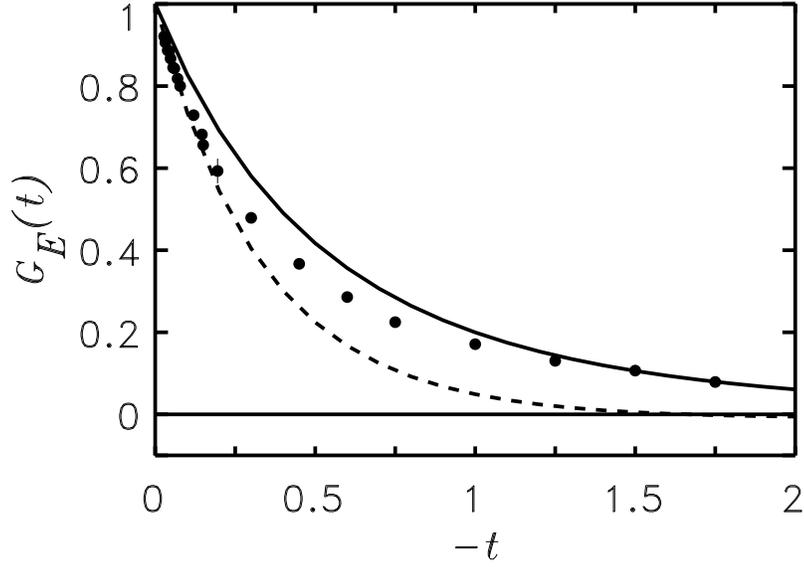,height=9cm}
\caption{$t$ dependence of the proton electric form factor $G_E(t)$
	for $\eta=0.35$ (solid) and $\eta=0.55$ (dashed).
	The bag radius in this and subsequent figures is fixed by
	$R M = 4 \omega_0$.
	The data are from Ref.\protect\cite{DATA1,DATA2}.}
\end{figure}

\begin{figure}
\label{fig3}
\epsfig{figure=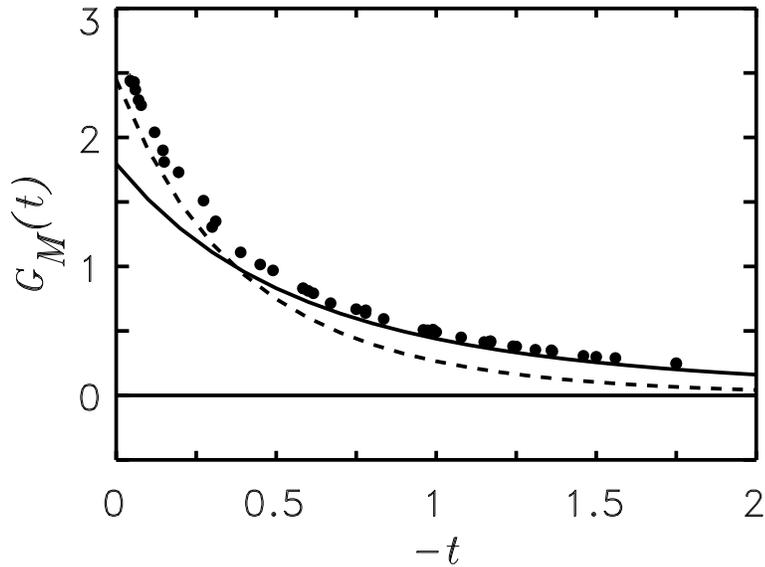,height=9cm}
\caption{$t$ dependence of the proton magnetic form factor $G_M(t)$
	for $\eta=0.35$ (solid) and $\eta=0.55$ (dashed).
	The data are from Ref.\protect\cite{DATA1,DATA2}.}
\end{figure}

\begin{figure}
\label{fig4}
\epsfig{figure=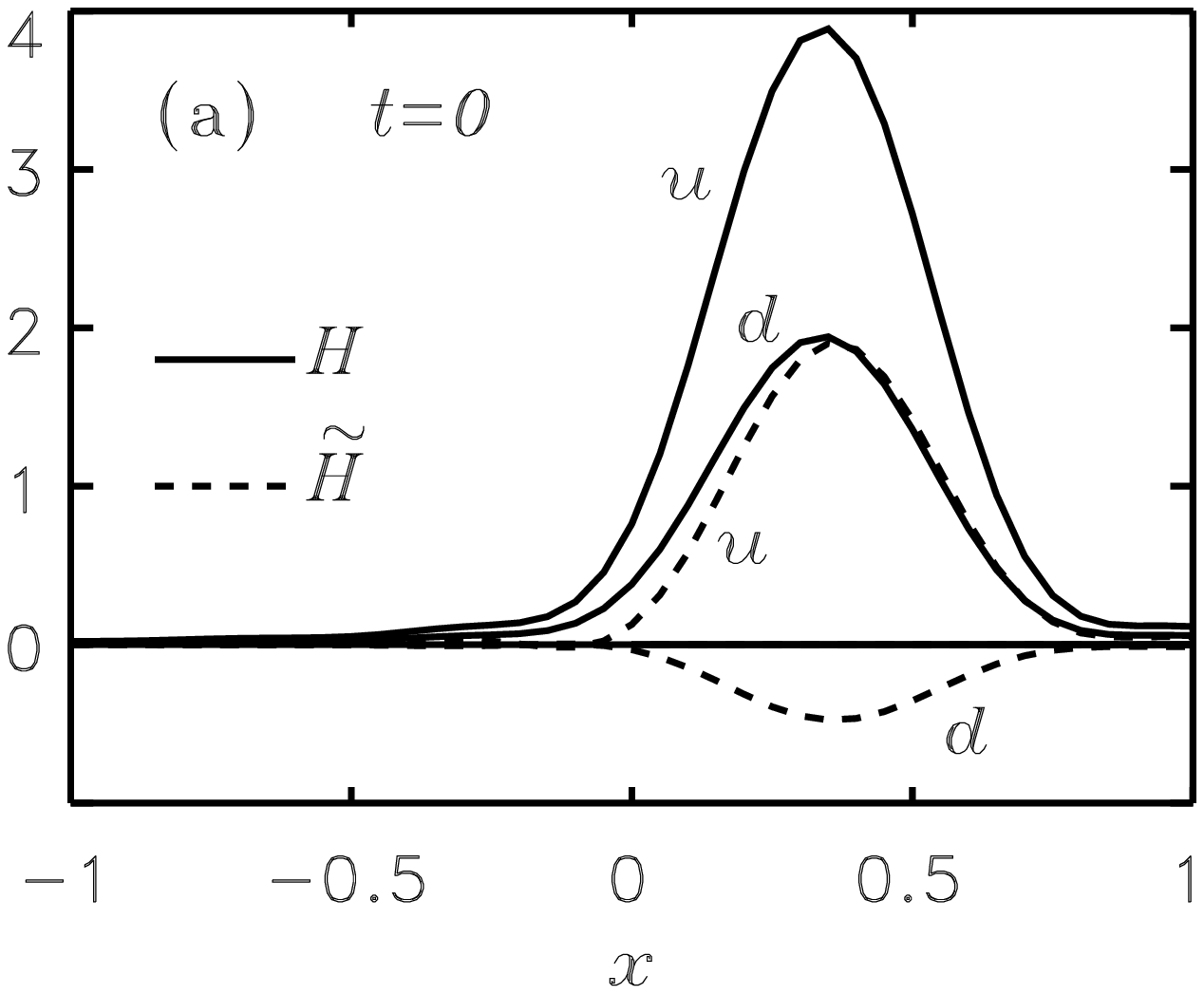,height=9cm}
\epsfig{figure=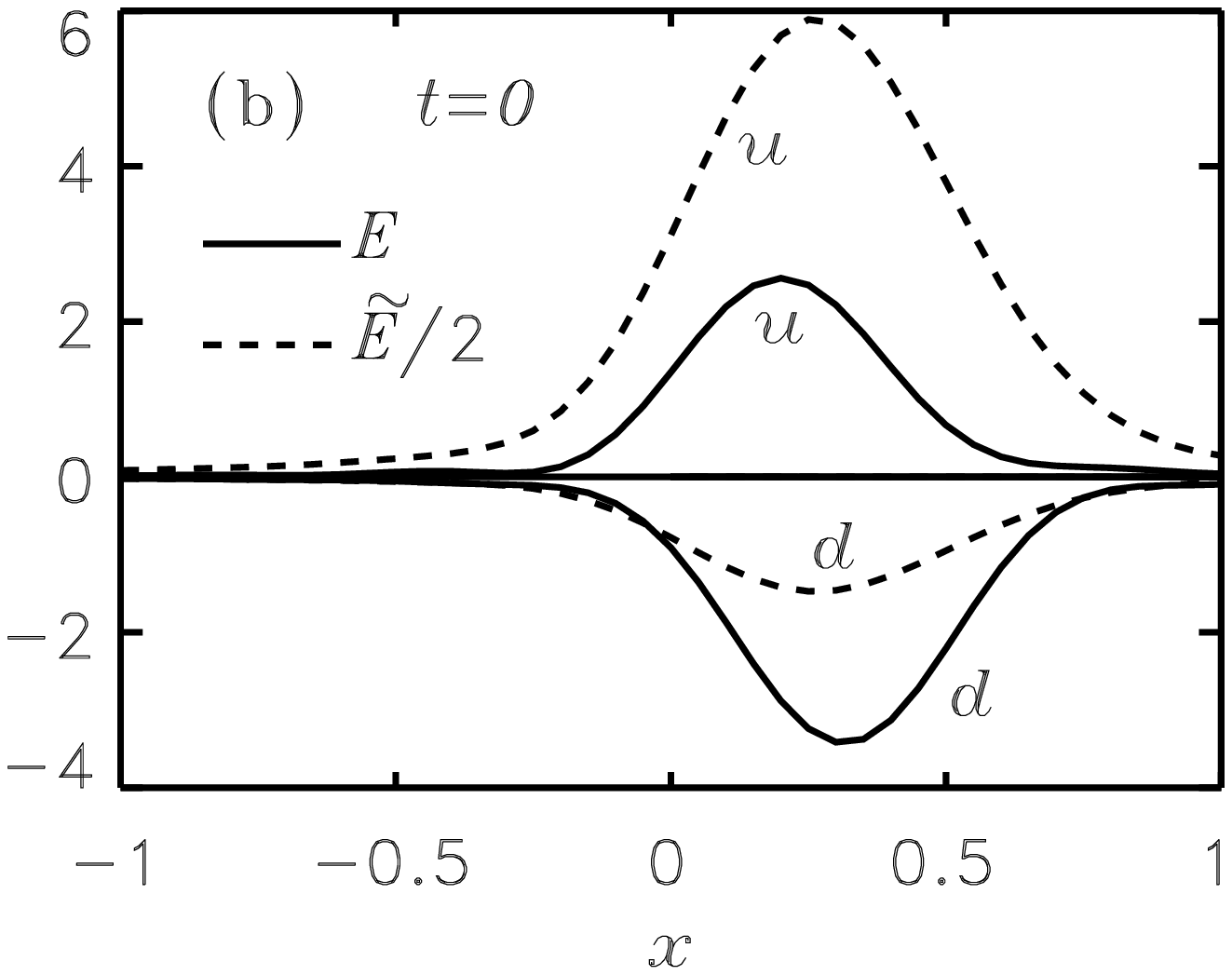,height=9cm}
\caption{Off-forward parton distributions at $t=0$ 
	for the $u$ and $d$ flavors:
	(a) $H$ and $\widetilde{H}$, which correspond
	to the usual spin averaged and spin dependent distributions
	$q(x)$ and $\Delta q(x)$, respectively;
	(b) $E$ and $\widetilde{E}$ for $\eta=0.55$.
	Note that $\widetilde{E}$/2 is plotted.}
\end{figure}

\begin{figure}
\label{fig5}
\epsfig{figure=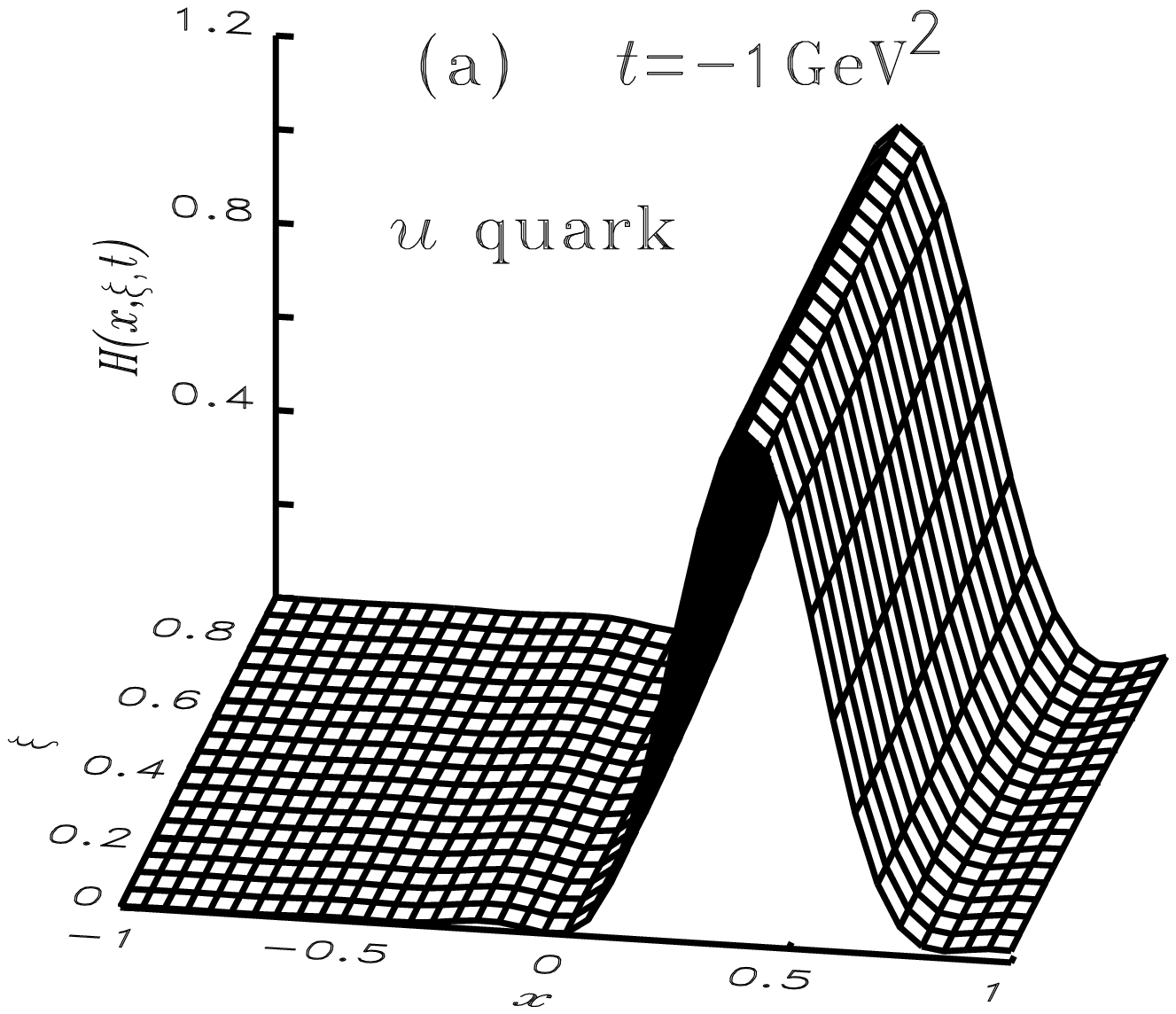,height=9cm}
\epsfig{figure=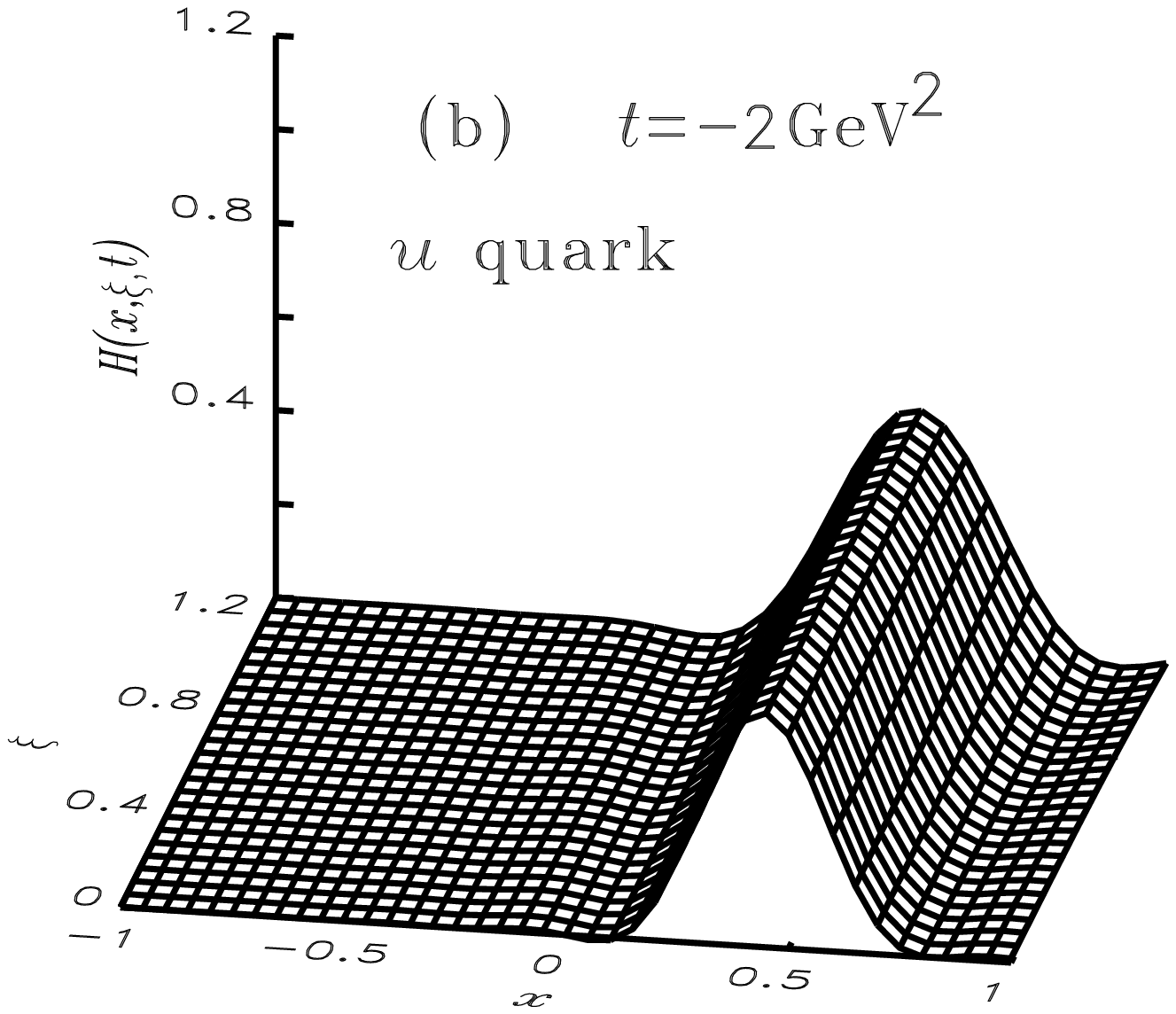,height=9cm}
\caption{Off-forward parton distribution $H(x,\xi,t)$ for the $u$ quark,
	as a function of $x$ and $\xi$, for 
	(a) $t = -1$ GeV$^2$ and (b) $t = -2$ GeV$^2$.}
\end{figure}

\begin{figure}
\label{fig6}
\epsfig{figure=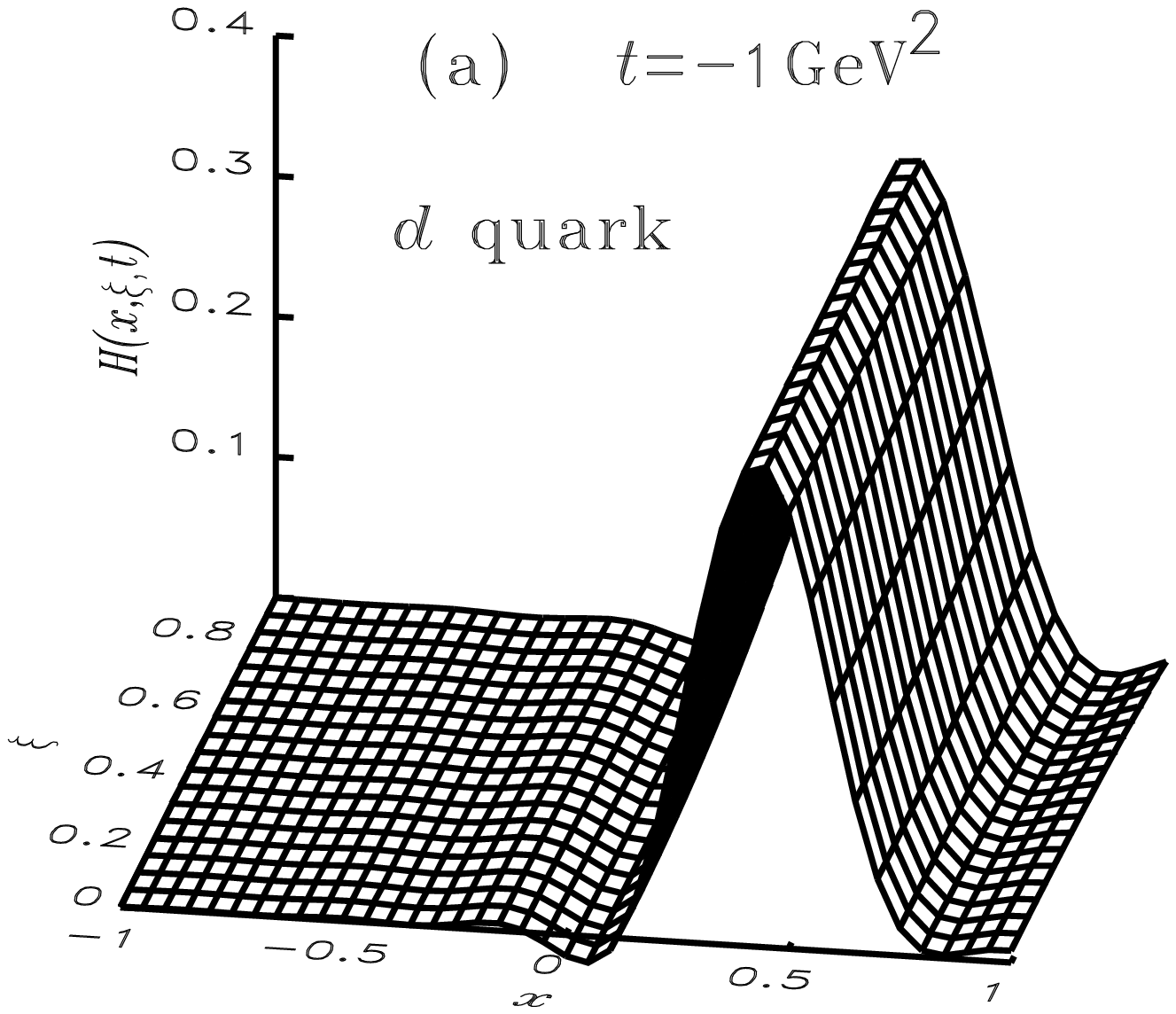,height=9cm}
\epsfig{figure=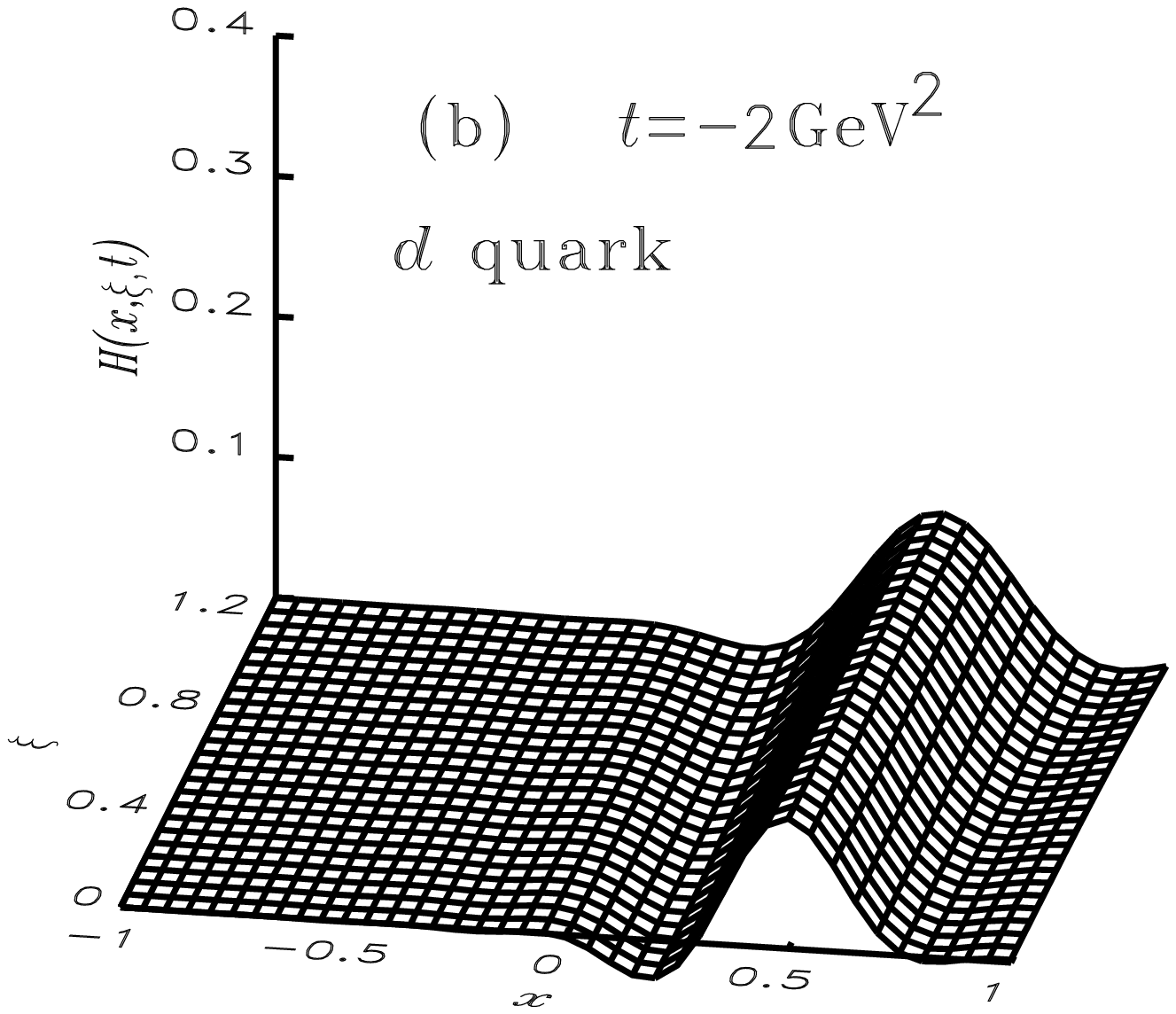,height=9cm}
\caption{As for Fig.~5, but for the $d$ quark.}
\end{figure}

\begin{figure}
\label{fig7}
\epsfig{figure=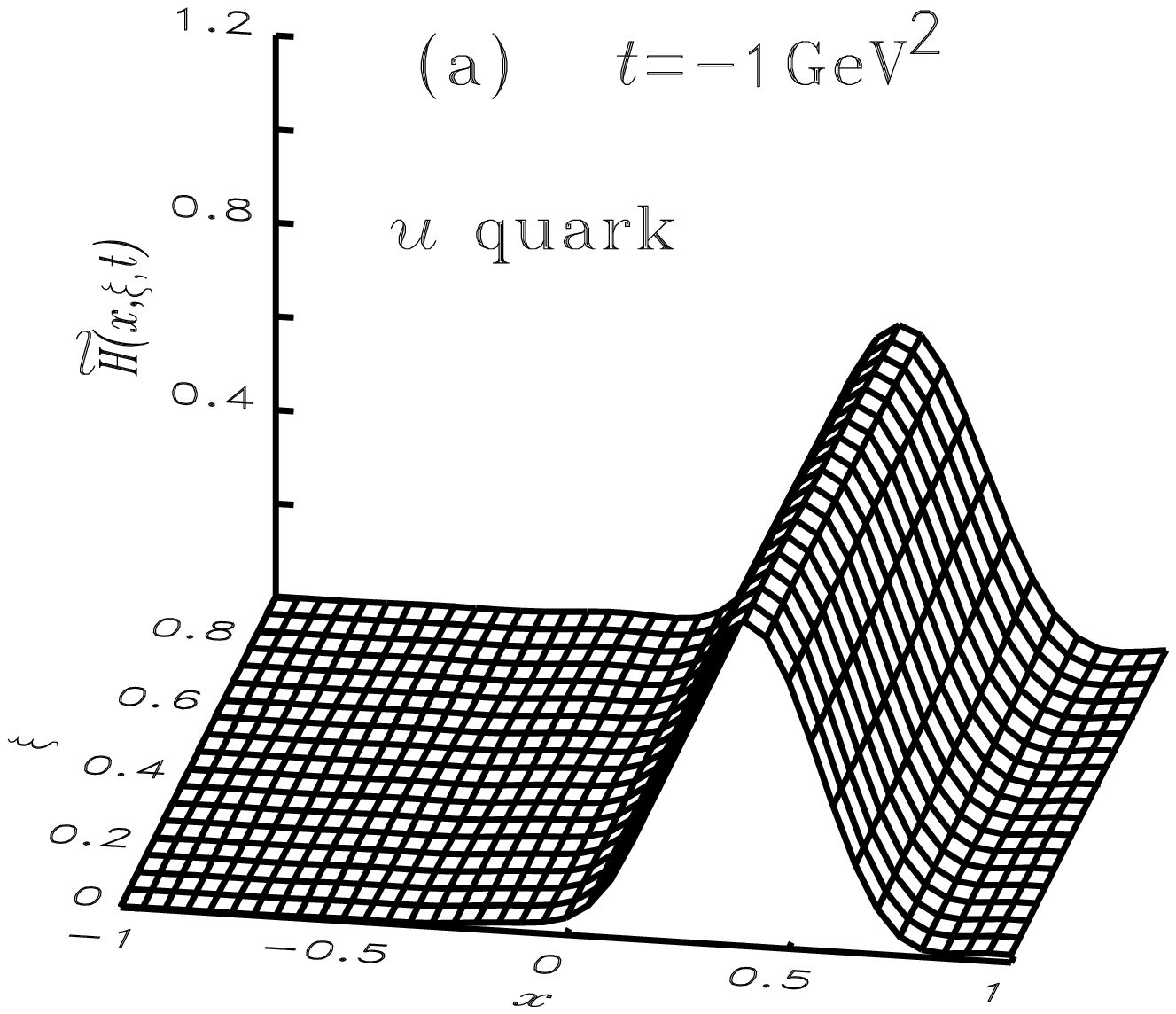,height=9cm}
\epsfig{figure=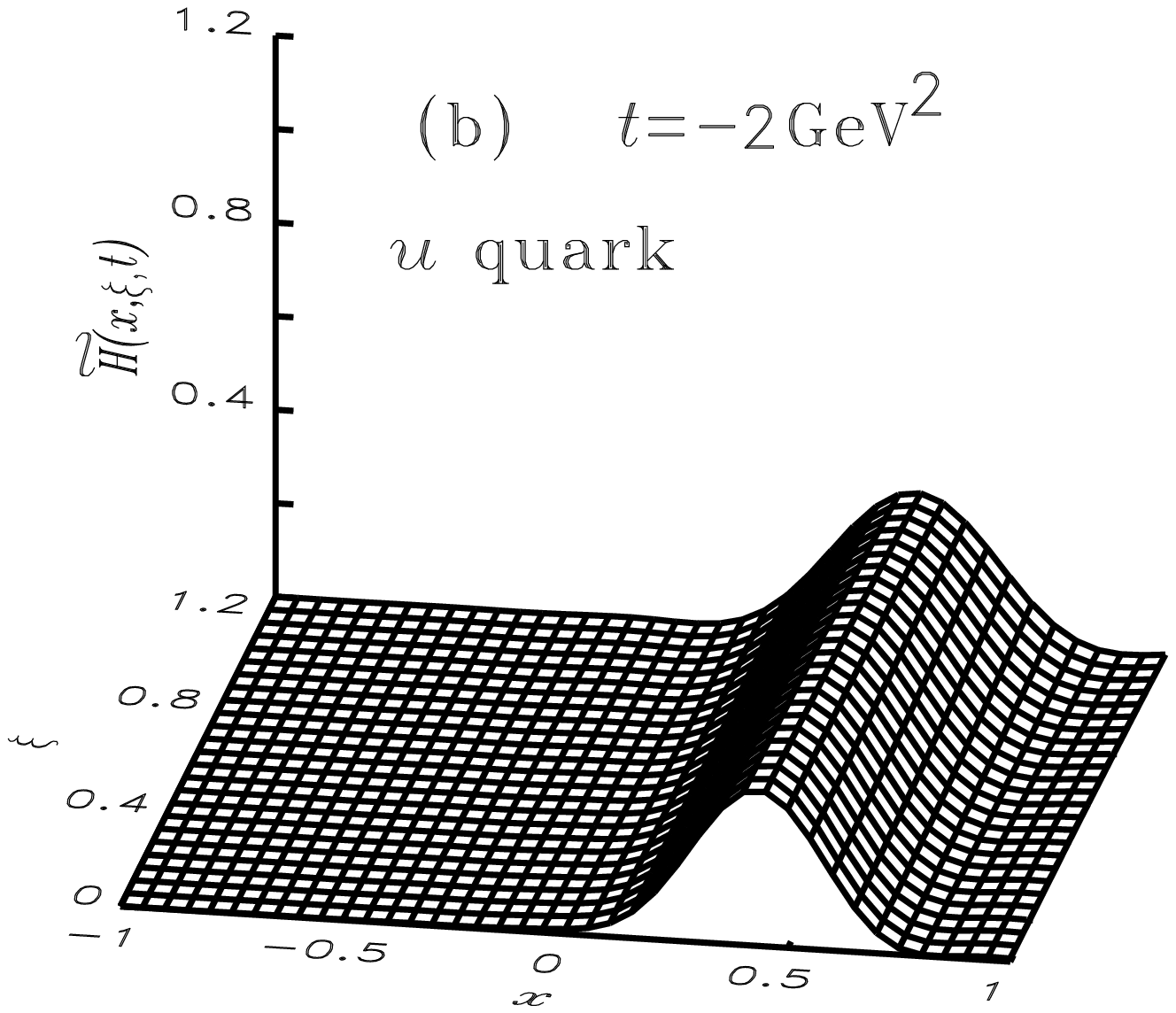,height=9cm}
\caption{Off-forward parton distribution $\widetilde{H}(x,\xi,t)$
	for the	$u$ quark, 
	for (a) $t = -1$ GeV$^2$ and (b) $t = -2$ GeV$^2$.
	The $d$ quark distribution is obtained by multiplying by -1/4.}
\end{figure}

\begin{figure}
\label{fig8}
\epsfig{figure=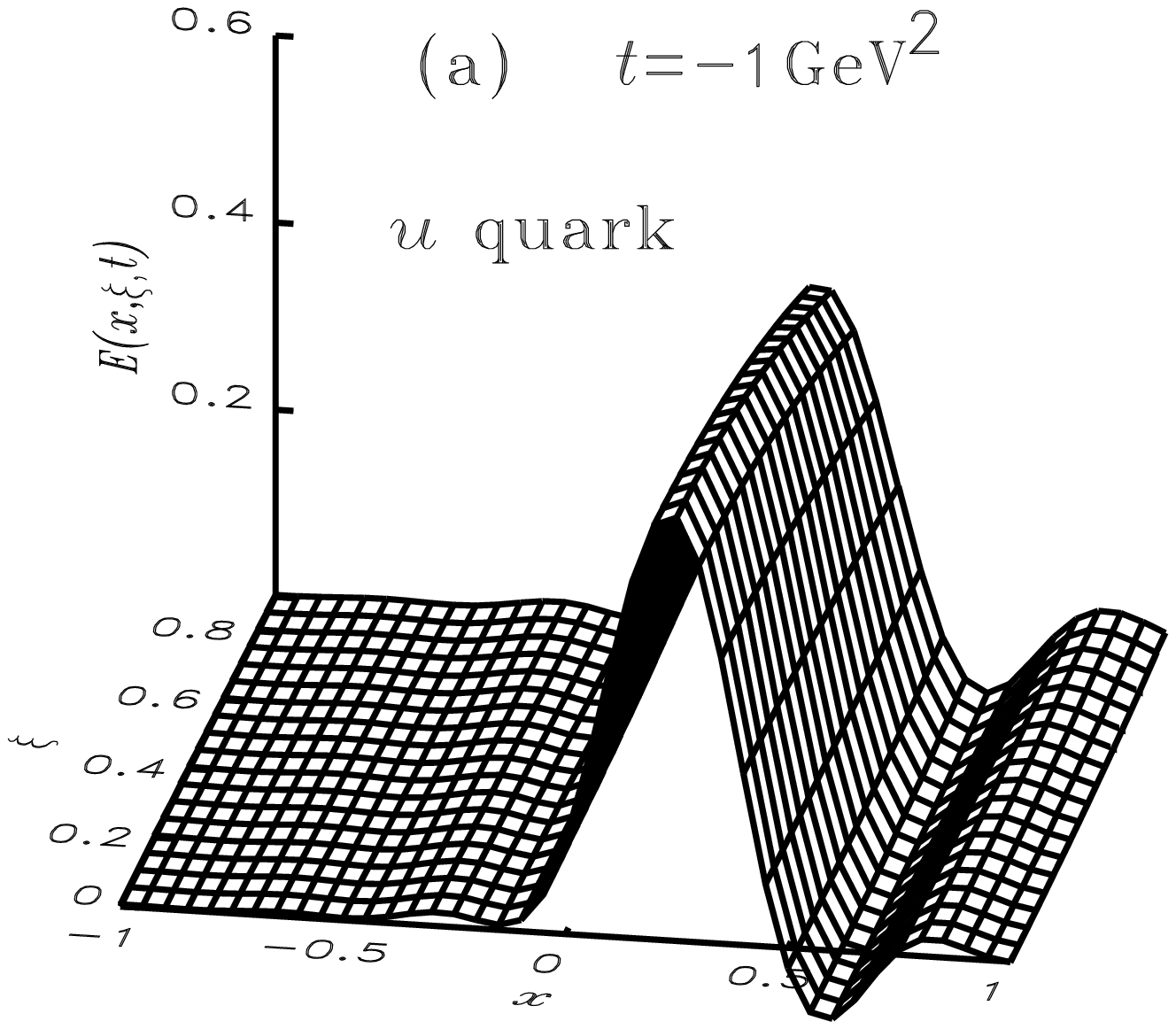,height=9cm}
\epsfig{figure=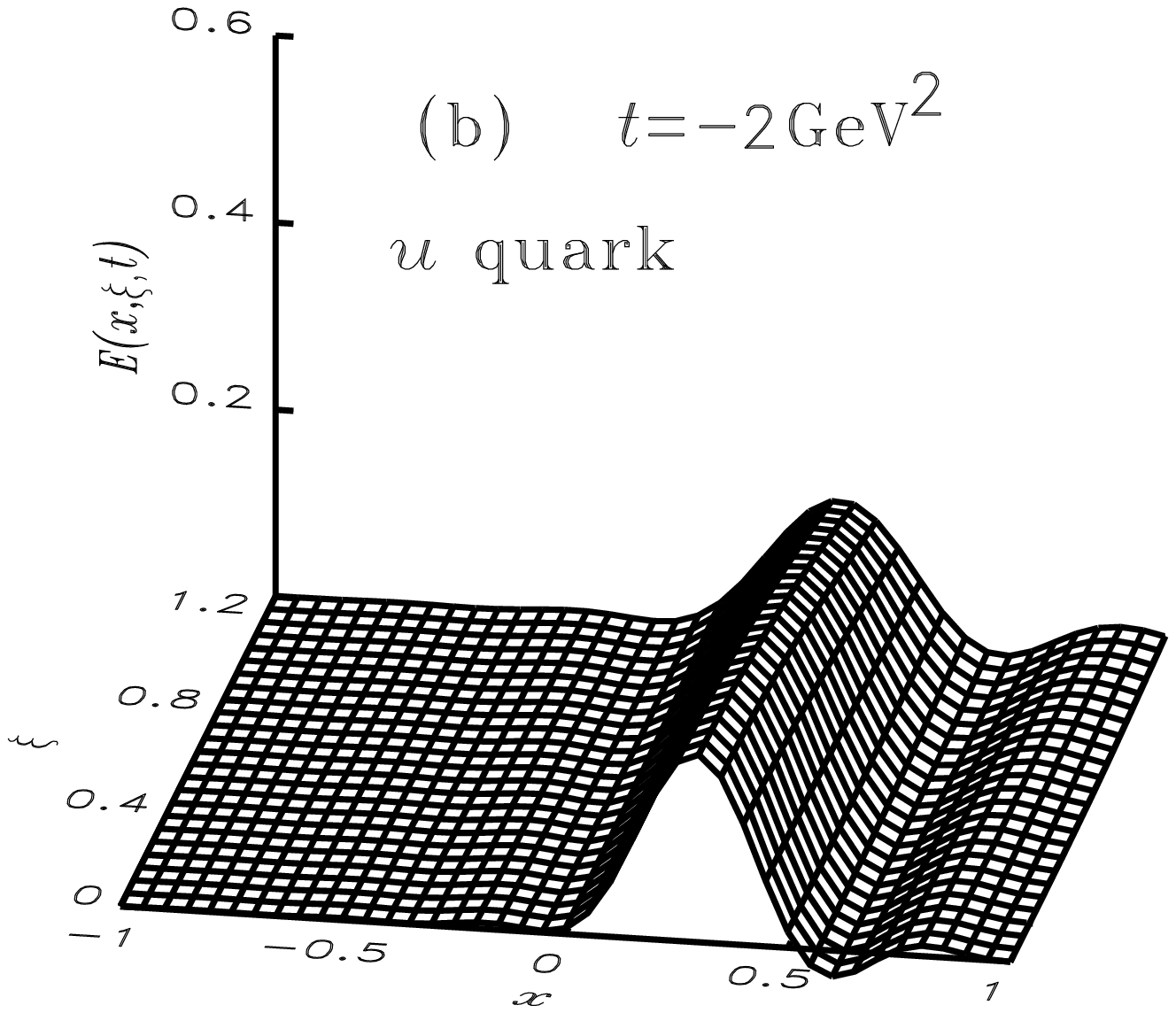,height=9cm}
\caption{Off-forward parton distribution $E(x,\xi,t)$ for the $u$ quark,
	for (a) $t = -1$ GeV$^2$ and (b) $t = -2$ GeV$^2$.}
\end{figure}

\begin{figure}
\label{fig9}
\epsfig{figure=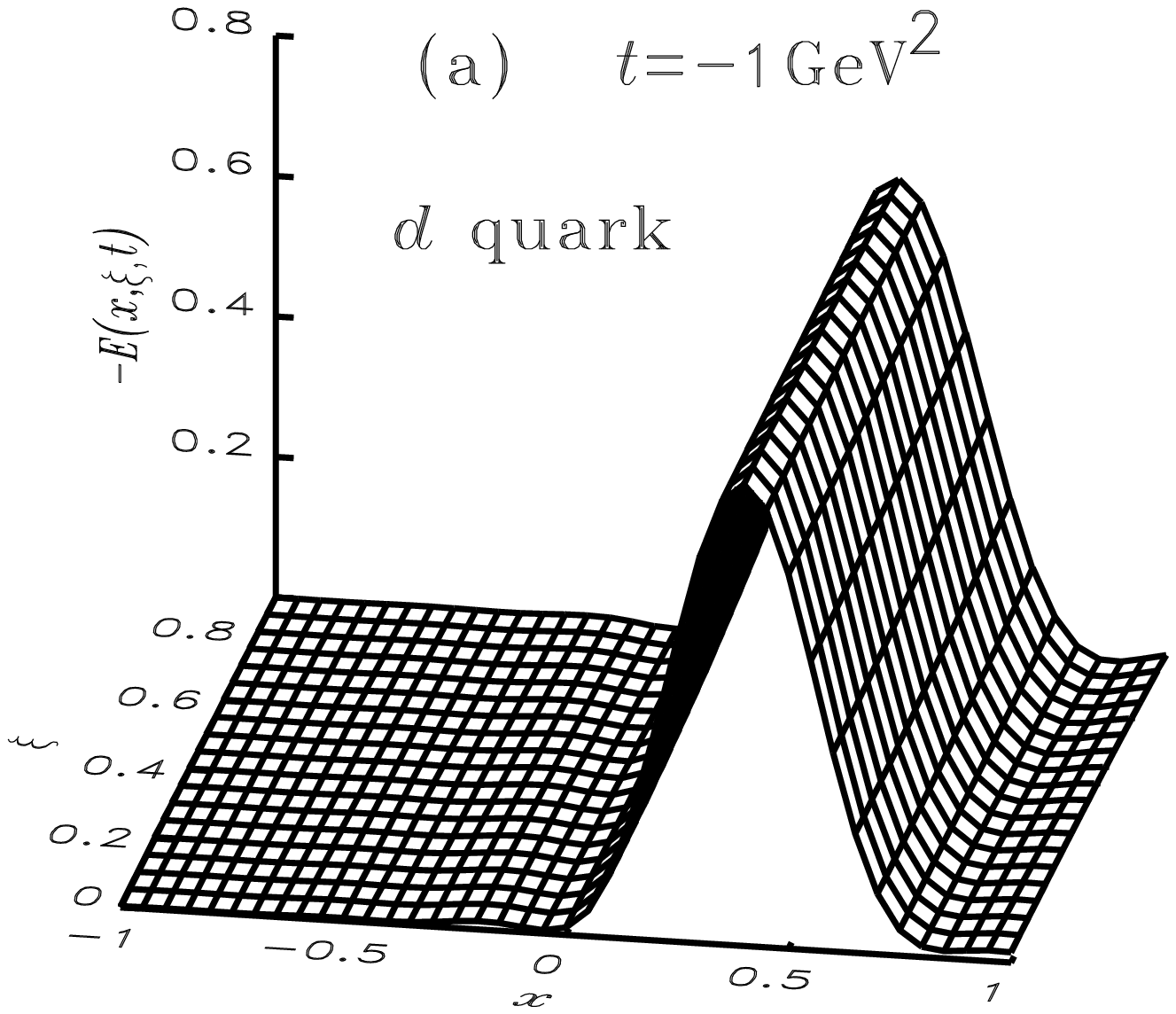,height=9cm}
\epsfig{figure=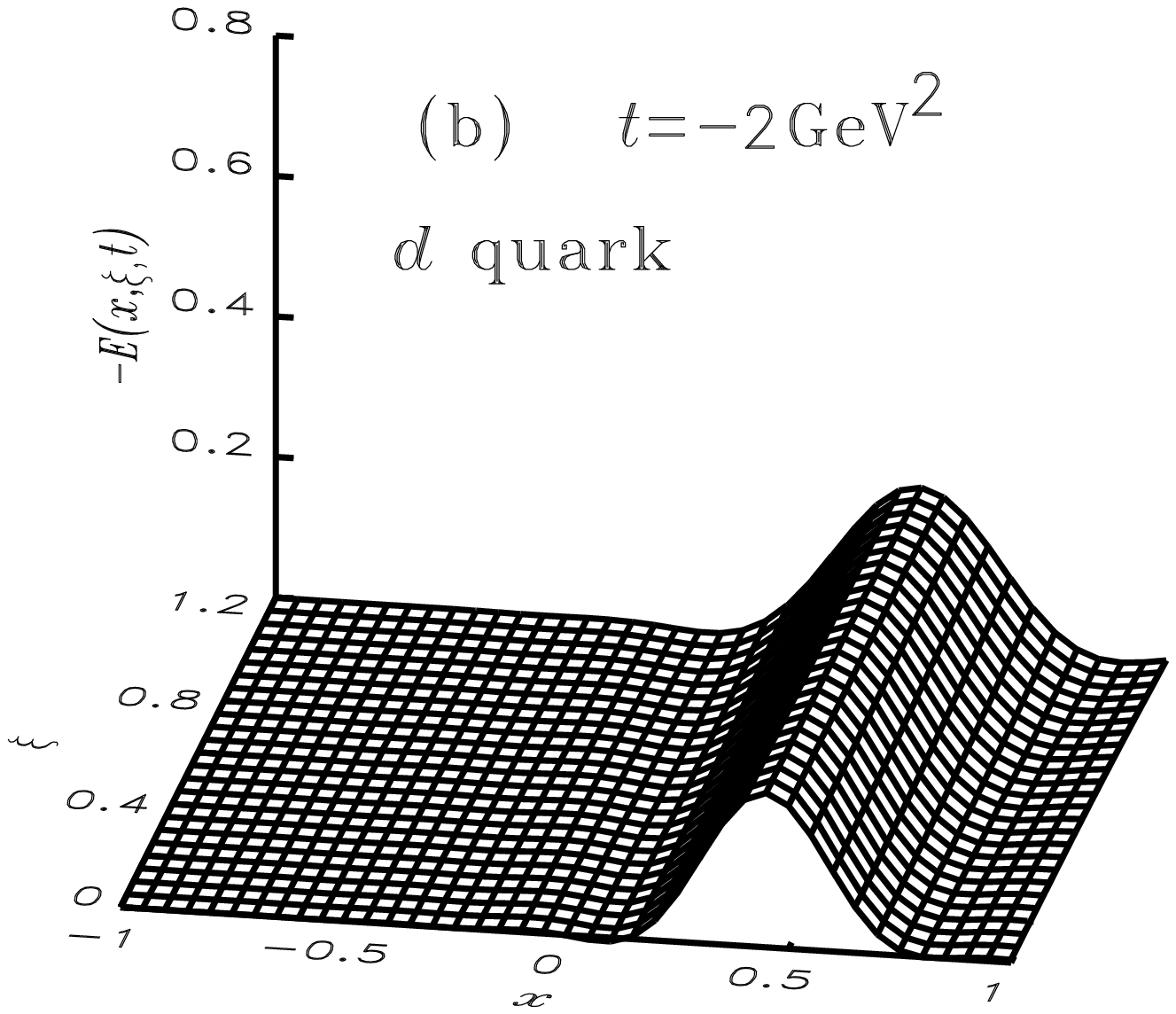,height=9cm}
\caption{As for Fig.~8, but for the $d$ quark.
	Note that it is $-E(x,\xi,t)$ which is plotted.}
\end{figure}

\begin{figure}
\label{fig10}
\epsfig{figure=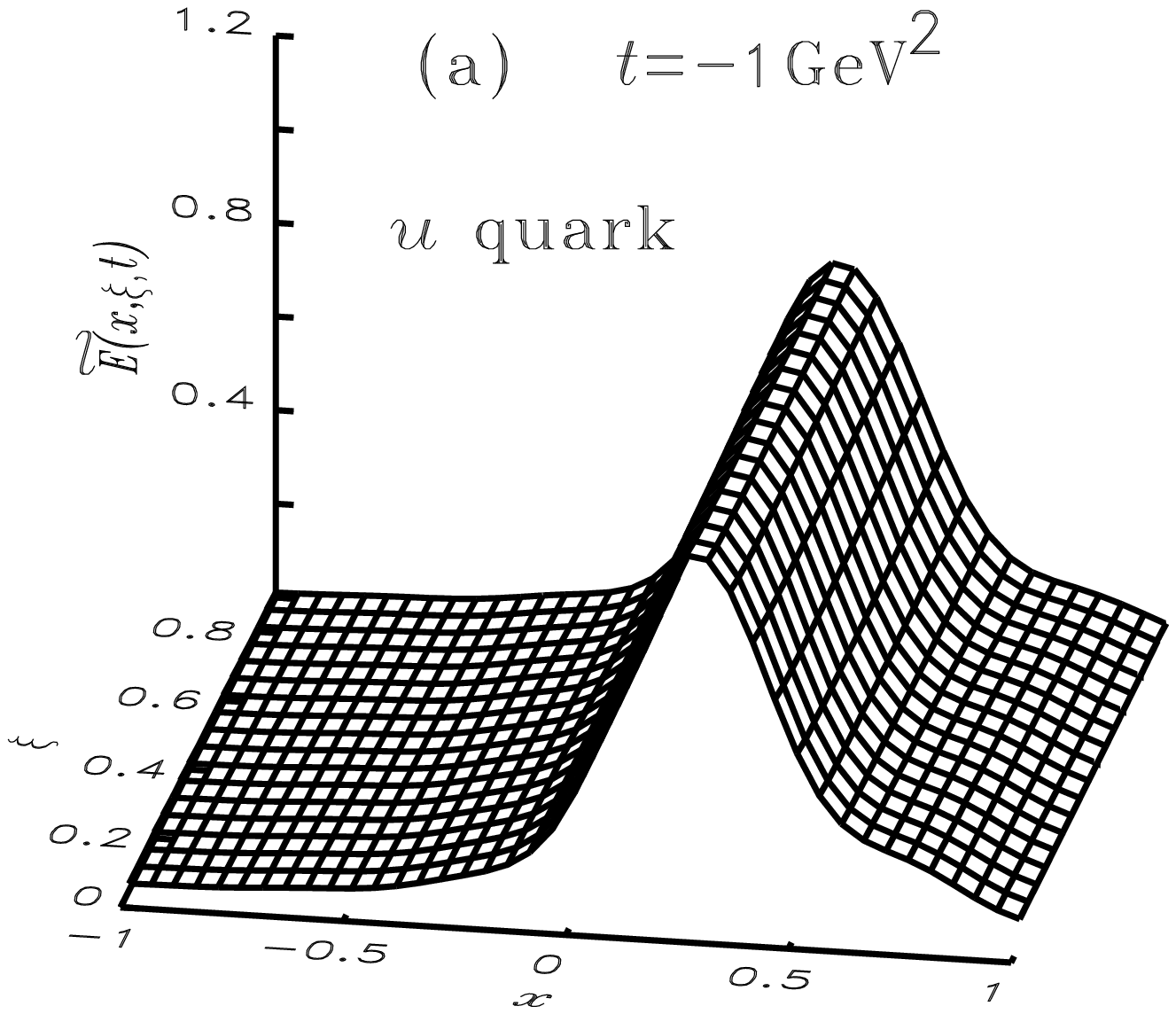,height=9cm}
\epsfig{figure=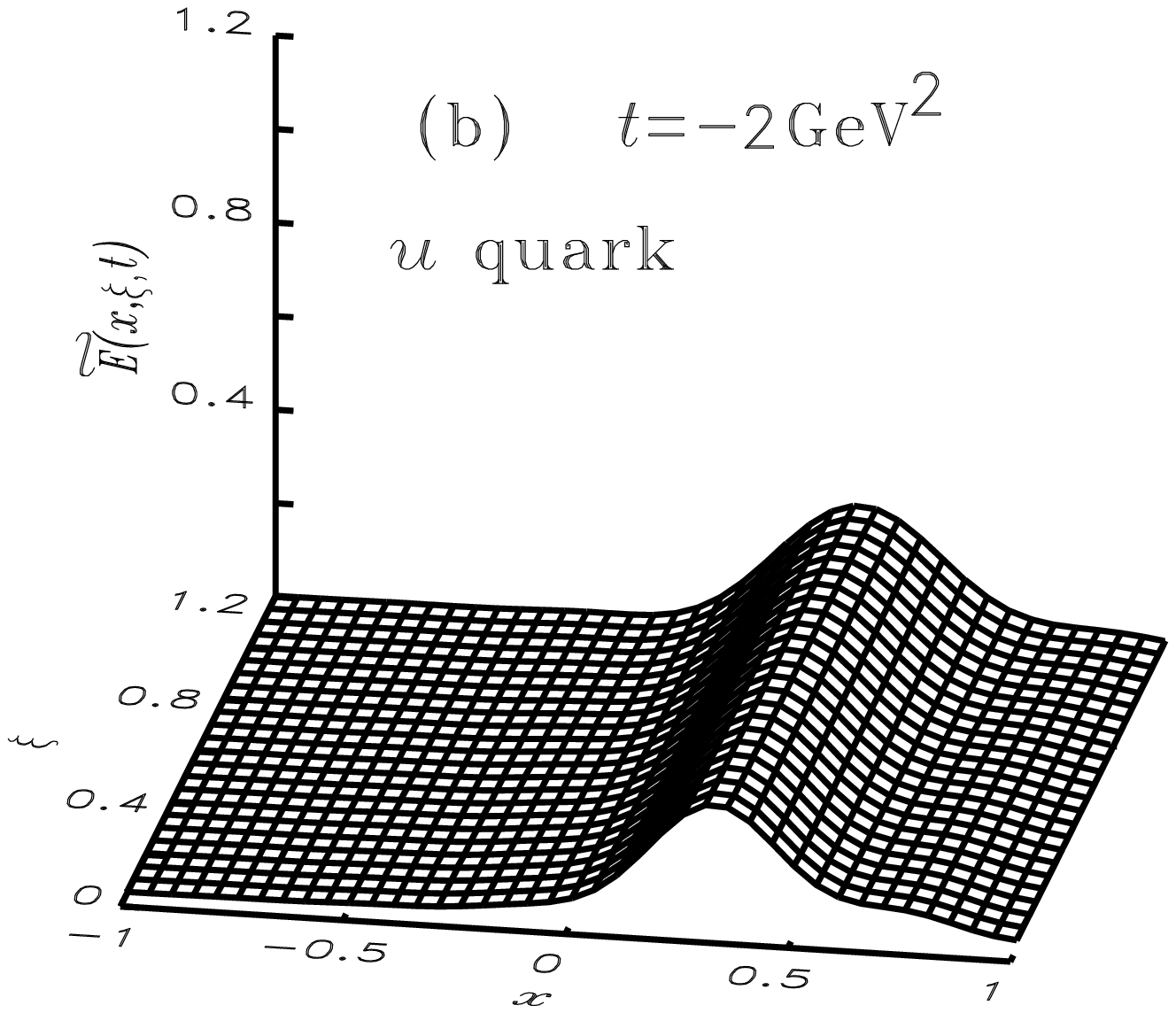,height=9cm}
\caption{Off-forward parton distribution $\widetilde{E}(x,\xi,t)$
	for the $u$ quark,
	for (a) $t = -1$ GeV$^2$ and (b) $t = -2$ GeV$^2$.
	The $d$ quark distribution is obtained by multiplying by -1/4.}
\end{figure}

\begin{figure}
\label{fig11}
\epsfig{figure=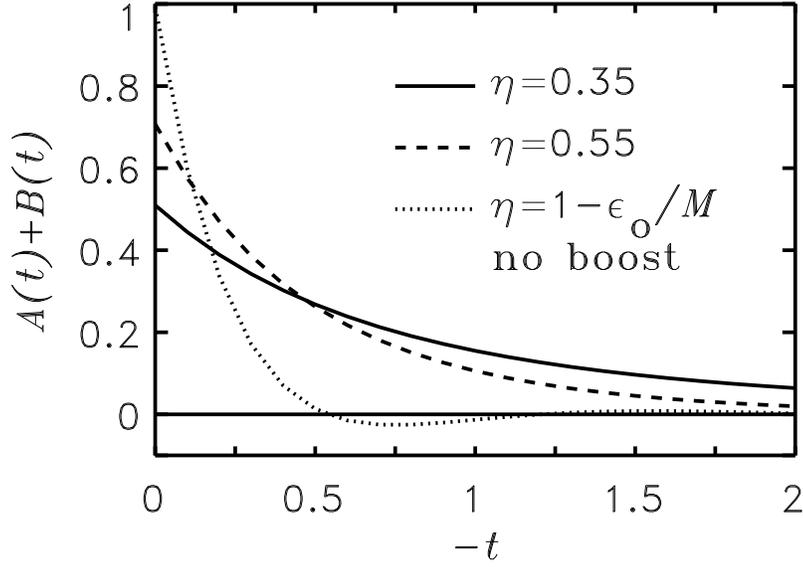,height=9cm}
\caption{$t$ dependence of the form factor $A+B$ of the nucleon 
	energy-momentum tensor, for the unboosted calculation 
	with $\eta=1-\epsilon_0/M$ (dotted), and for the 
	boosted result with $\eta=0.55$ (dashed)
	and $\eta=0.35$ (solid).}
\end{figure}

\begin{figure}
\label{fig12}
\epsfig{figure=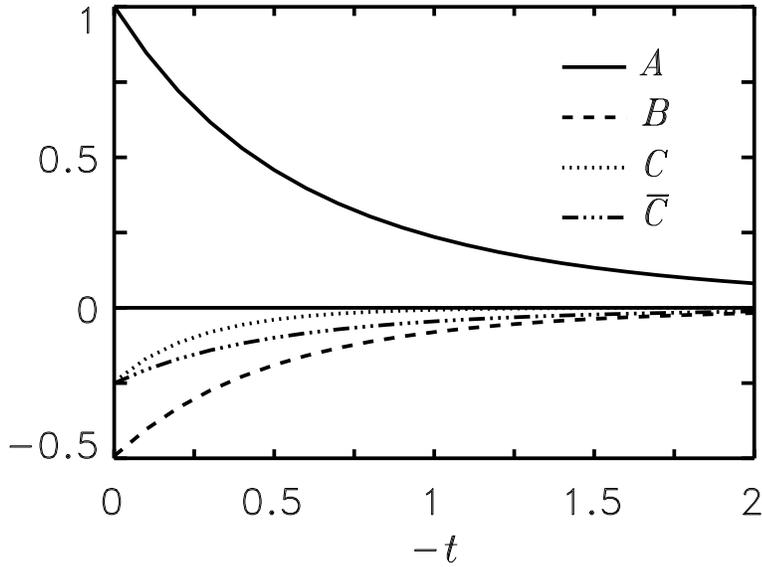,height=9cm}
\caption{$t$ dependence of the form factors $A, B, C$ and $\overline C$
	of the energy-momentum tensor, for $\eta=0.35$.}
\end{figure}

\end{document}